\documentstyle[epsf]{article}

\newcommand{\bee}{\begin{equation}}
\newcommand{\ee}{\end{equation}}
\newcommand{\beea}{\begin{eqnarray}}
\newcommand{\eea}{\end{eqnarray}}

\newcommand{\ewxy}[2]{\setlength{\epsfxsize}{#2}\epsfbox[10 60 640 570]{#1}}

\begin{document}
\thispagestyle{empty}
\parskip=12pt
\raggedbottom

\def\mytoday#1{{ } \ifcase\month \or
 January\or February\or March\or April\or May\or June\or
 July\or August\or September\or October\or November\or December\fi
%\space\number\day ,
 \space \number\year}
\noindent
\hspace*{9cm} COLO-HEP-410\\
\vspace*{1cm}
\begin{center}
{\LARGE Optimizing the Chiral Properties of Lattice Fermion Actions}

\vspace{0.5cm}

Thomas DeGrand,
Anna Hasenfratz,  and Tam\'as G.\ Kov\'acs\\
Physics Department, 
        University of Colorado, \\ 
        Boulder, CO 80309 USA

(the MILC collaboration)

\begin{abstract}
We describe a way to optimize the chiral behavior  of 
Wilson-type lattice fermion actions by studying the low energy real 
eigenmodes of the Dirac operator.
We find a candidate action, the clover action with fat links with a tuned
 clover term.
We present a calculation of spectroscopy and matrix elements at
Wilson gauge coupling $\beta=5.7$. The action shows good scaling behavior.
\end{abstract}

\end{center}
\eject

\section{Introduction}
There is an increasing accumulation of evidence from lattice simulations
of the importance of the topological
properties of the QCD vacuum \cite{VACUUM}.  The (lattice) pure gauge
QCD vacuum is filled with instantons of average radius $\sim 0.3$ fm,
with a density of about 1 fm$^{-4}$, and they are responsible for a large
part of chiral symmetry breaking in QCD \cite{IILM,SU3INST}.
The extent to which the complete continuum phenomenology of instanton effects
in hadronic physics is
actually realized by QCD, and can be studied using lattice simulations,
remains an open question, but its outline is present.

Lattice artifacts associated with instantons also seem to be connected to
some of the difficulties of numerical simulations with quenched QCD, 
namely, eigenmodes of the Dirac operator
which occur away from zero bare quark mass, and which spoil the
calculation of the fermion propagator, 
the so-called ``exceptional configurations'' \cite{FNALINST,NEW_EXCEPT}.

In this work we wish to look at topology and fermions on the lattice
from a different perspective:  given that instantons
are responsible for chiral symmetry breaking, is it possible to optimize
the lattice discretization of a fermion action with respect to its topological
properties? Our   goal is to shrink the range of bare quark mass
over which  the low lying real eigenmodes of the Dirac operator occur,  on background gauge field configurations
which are typical equilibrium configurations
at some gauge coupling.

The action we propose is the standard clover action, except that
the gauge connections are replaced by APE-blocked \cite{APEBlock}
 links, and the
clover coefficient is tuned to optimize chiral properties.
Specifically,
\beea
S & = &\sum_n (m+4)\bar \psi(n) \psi(n) \nonumber  \\
& - & {1\over 2}  \sum_{n \mu}(\bar \psi(n)
(1 - \gamma_\mu) V_\mu(n) \psi(n+ \mu) 
+
\bar \psi(n)(1 + \gamma_\mu) V_\mu^\dagger(n-\mu) \psi(n - \mu)
\nonumber  \\
& + & {C \over 2} \sum_{n,\mu,\nu} \bar \psi(n) i \sigma_{\mu\nu} 
P_{\mu\nu}(n) \psi(n)
\label{FATCLOVER}
\eea
with  
\beea
V^{(n)}_\mu(x) = &
(1-c)V^{(n-1)}_\mu(x) \nonumber  \\
& +   c/6 \sum_{\nu \ne \mu}
(V^{(n-1)}_\nu(x)V^{(n-1)}_\mu(x+\hat \nu)V^{(n-1)}_\nu(x+\hat \mu)^\dagger
\nonumber  \\
& +  V^{(n-1)}_\nu(x- \hat \nu)^\dagger
 V^{(n-1)}_\mu(x- \hat \nu)V^{(n-1)}_\nu(x - \hat \nu +\hat \mu) ),
\label{APE}
\eea
with  $V^{(n)}_\mu(x)$  projected back onto $SU(3)$ and
 $V^{(0)}_\mu(n)=U_\mu(n)$ the original link variable. 
We take $c=0.45$ and $N=10$ smearing steps,
chosen because of our previous work in instantons \cite{SU3INST}.
This choice of parameters is not unique and might not even be optimal.
$P_{\mu\nu}$ is the usual clover set of links,
but built of the $V_\mu$'s. At Wilson gauge coupling $5.7-5.8$, the
best choice is $C=1.2$, and it decreases to the tree-level $C=1$ value
at larger $\beta$.
We will also quote simulation data in terms of
 the hopping parameter, ${1 \over 2 \kappa} = m + 4$.

The optimized  action has a spread of  low lying real eigenmodes with respect to the
bare quark mass which is less then a third  of the usual Wilson action.
In terms of the square if the pion mass the spread is about three times
smaller for the optimized action than
for the standard Wilson action.
The action has other good features, as well: the renormalization factors connecting
lattice quantities to their continuum values appear to be very close
to unity.  The action also appears to require only about half the
number of sparse matrix inversion steps as the usual clover action 
(at equivalent values of physical parameters).

It is possible to tune the standard clover action (with the original links
used as connections) to improve the interaction of quarks with instantons.  
This choice is not the best one. It happens that the
best value of $C$  for $\beta<6.0$ is large enough that a new class
of exceptional configurations appear.
These configurations compromise simulations.
The connection between exceptional configurations and instantons is
explored in detail in a companion work by us \cite{NEW_EXCEPT}.

The action we propose is completely unimproved in its kinetic properties,
so its dispersion relation and heavy quark mass artifacts are identical to
those of the Wilson action.  It would be very easy to improve it by
beginning with a more complicated free fermion discretization.

Fermion actions with fat links have received considerable attention
in the past year.
The first use of them we know of (although  with
a  different motivation) was in the simulations of QCD on
cooled gauge fields by the MIT group \cite{MITCOOL}.
The MILC collaboration \cite {MILC}, Orginos and Toussaint \cite{ORT}
and 
Laga\"e and D.~K. Sinclair\cite{SINCLAIR} have shown that modest fattening
considerably improves flavor symmetry restoration in simulations with staggered
fermions.
We have used calculations of staggered spectroscopy on highly smoothed
gauge configurations
to compare chiral symmetry breaking  in 
$SU(2)$ gauge theory and in instanton backgrounds \cite{COLOINST}.
Finally, all fixed point actions and approximate fixed point  actions
we know of \cite{ALLFP,HYPER}
for fermions seem to incorporate fat links.
Fixed point fermions realize the index theorem and retain
chiral symmetry at nonzero lattice spacing \cite{PLN}, and so one
way of viewing a fat link action is as an approximate FP action,  which
includes its chiral properties but not its kinetic ones.

An apparent drawback of fat link actions is the lack of a transfer matrix
between consecutive time slices. N APE steps can mix links up to $\pm N$
timeslices away and a strict transfer matrix cannot be defined on time slices
closer than $2N$ lattice spacings (20 in our case).
In practice we are concerned only with the exponential decay of correlation
functions and we found asymptotic decay after 3-5 time slices in our
simulation. This is expected if we realize that APE smearing is basically a
random walk whose range can be estimated as $\sqrt{N} c\approx 1.4$.

The outline of the paper is as follows: In Section 2 we review the
continuum index theorem and describe how lattice fermions fail to
reproduce it. We then describe how we find the real eigenmodes
of the Wilson-Dirac operator.
All this is basically a review, and experts may skip it.
In Section 3 we describe the tests we performed to tune the action.
Section 4 describes a calculation of spectroscopy and matrix elements 
at $aT_c=1/4$ (Wilson gauge coupling $\beta=5.7$) using the new action.

\section{Measuring the Chiral Properties of Wilson-like Fermions}

The reader might recall that the local topological density
\bee
q(x) = {1\over {32\pi^2}}\epsilon_{\mu\nu\rho\sigma}F_{\mu\nu}F_{\rho\sigma}
\ee
is related to the divergence of the flavor singlet (with $n_f$ flavors)
 axial-vector current  
\bee
\partial_\mu \bar \psi i \gamma_\mu \gamma_5 \psi =
 2 m \bar \psi i \gamma_5 \psi + 2in_f q.
\ee
Integrating over all $x$, this relation implies
 a connection between the topological charge $Q=\int d^4x q(x)$
and a mode sum,
\bee
Q = m {\rm Tr} \bar \psi \gamma_5 \psi =
 m {\rm Tr} \gamma_5 {1\over{ \gamma \cdot D + m}} = m \sum_s {{f^\dagger_s \gamma_5 f_s}\over{i \lambda_s + m}}
\ee
where $f_s$ are the eigenfunctions of the (antihermetian) Dirac operator
$\gamma \cdot D$, $\gamma \cdot D f_s = i \lambda_s f_s$, with $\lambda_s$ real.
The property $\{\gamma_5 , \gamma \cdot D\}=0$ leads to the condition that
$f_s^\dagger \gamma_5 f_s =0$ if $\lambda_s \ne 0$, and if $\lambda_s=0$
we may choose $f_s$ to have a definite chirality, $\gamma_5 f_s = \pm f_s$.
Thus it follows that
\bee
Q = \sum_{s, \lambda_s=0}  f_s^\dagger \gamma_5 f_s = n_+ - n_-
\ee
where $n_+$ and $n_-$ are the number of zero eigenmodes with positive and negative
chirality. This is the index theorem.

On the lattice, essentially every statement in the preceding paragraph
 is contaminated by lattice artifacts \cite{SMITVINK}. 
Here we focus on Wilson-like
fermions, where chiral symmetry is broken by the addition of terms
proportional to the Dirac scalar and/or tensor operators.
The Wilson  or clover fermion
action analog of $\gamma \cdot D$, $D_w$, is neither
Hermetian nor antihermetian and its eigenvalues are generally complex.
$D_w$ can also have real eigenvalues. These real eigenvalues usually do not occur
 at zero bare quark mass.
Their locations spread across a range of quark mass values.

On smooth, isolated instanton background configurations,
the location of the real eigenmode varies with the size $\rho$ of the instanton.
For large instantons, the low lying real mode occurs close to zero quark mass.
Accompanying these near-zero modes are a set of modes which 
do not cluster around $m_0=0$, but around
$-am_0 \simeq O(1)$.  These are ``doubler modes.''  
As one decreases the instanton size, the eigenmodes at $am_0 \simeq 0$ shift towards negative quark
masses, approaching the doubler modes. 
This $\rho$-dependent mass shift and the 
location of the doubler modes all depend on the particular choice of lattice
fermion action \cite{NEW_EXCEPT,SIMMA,SCRI_smooth}.
The shift of the low lying real eigenmode for a given instanton size for
the Wilson action is larger than for the clover
action, which has better chiral properties. Nevertheless, for both actions,
as the instanton size decreases,  sooner or later the low energy eigenmode
shifts to  large negative quark mass,
approaches the doublers and eventually annihilates with one of the doublers
- the fermion does not see the
instanton any longer.

On equilibrium background configurations, the shift of the eigenmode
with $\rho$ is accompanied by an overall $\beta$-dependent mass shift.
On top of that there is the spread of the eigenmodes according to the sizes of the background
instantons.
The spread of these low energy modes, if distinguishable from the doublers, 
characterize the amount of explicit chiral symmetry
breaking of the fermionic action. 
On configurations with small lattice spacing the typical instanton is large in
lattice units. The spread of the low energy modes is small and they are well separated from the
doublers. The explicit chiral symmetry
breaking of the fermionic action is small and controlled. 
On configurations with large lattice spacing the instantons are small (if present at all). The spread
of the low lying modes is large, close to or overlapping with the doubler modes.
The hadron spectrum on these configurations could numerically be similar to the continuum spectrum, 
but the physical mechanism behind it is very different from continuum QCD.

Determination of the locations of the real roots of $D_w +m$ in a particular 
background gauge configuration can be done very simply.
We approximate
\bee
P(m_0) = \langle \bar \psi \gamma_5 \psi\rangle
 = {\rm Tr} (D_w+m_0)^{-1}\gamma_5
\ee
with a noisy estimator: cast a random vector $\eta_i$ on each site
$i$, compute $(D_w+m_0) \chi = \eta$ and measure
\bee
P(m_0)_\eta = \sum_{ij} {\rm Tr} \bar \eta_i \gamma_5 \chi_i
\label{PG5P}
\ee
as a function of $m_0$.
The eigenmode of $D_w$ is located at $-m_0$ and its appearance is signaled
by the appearance of a pole in $P(m_0)$.
This is a variant on the standard method of measuring $\langle \bar \psi \psi
\rangle$ in a lattice simulation. It has also seen considerable use by the Fermilab
group \cite{FNALINST}
in their studies of exceptional configurations (they use a flat source
$\eta_i=$ constant, not a noisy source).
The eigenfunction itself of a particular
mode can be found by performing the inversion for a test mass very close
to an eigenvalue. Then 
\bee
\chi = \sum_s {{ f_s (f_s^\dagger,\eta) }\over {\lambda_s + m_0} }
\ee
is saturated by the mode closest to the pole, and $\chi^\dagger \chi$
is (proportional to) the probability density for that mode.

\section{Testing and Tuning Actions}

We study the locations of low energy real eigenmodes in equilibrium (quenched)
gauge configurations. Here there are two issues: First, do the
low lying modes separate from the doubler modes, and second, what
is the spread of the low lying modes.

The first question is important because, if the low lying modes do
not separate from the doubler modes, then the physics of chiral symmetry
breaking with a lattice cutoff is different than in the continuum.
The lattice theory is no longer just a crude approximation to the continuum,
and one cannot speak of chiral symmetry breaking as being induced by
instantons.  Chiral symmetry is certainly broken in QCD for
any value of the cutoff, including the strong coupling limit \cite{STRONG},
but the mechanism does not involve instantons.

Note that if the low lying modes do not separate from the 
doubler modes, it does not make sense to talk about the spread of
the low lying modes.

In this paper we restrict ourselves to clover-like actions.
Our goal is to tune the clover coefficient for good chiral behavior, i.e. to minimize the spread of the
low lying modes. We could attempt this 
program on the original configurations. However, it is  known that rough
gauge configurations with large clover coefficient are plagued by exceptional configurations. This
problem is greatly reduced if the links of the configurations are smoothed by a series of APE smearing
steps \cite{NEW_EXCEPT}. We also know from previous work \cite{COLOINST} that an action with fat
links is insensitive to  short distance fluctuations, but still knows about instantons and the
additional long
distance behavior of the gauge
field responsible for confinement. So we will begin with a fat link action of Eqn. \ref{FATCLOVER}.
Throughout this paper we create the fat link by 10 APE smearing steps with smearing coefficient
$c=0.45$.

\begin{figure}
\centerline{\ewxy{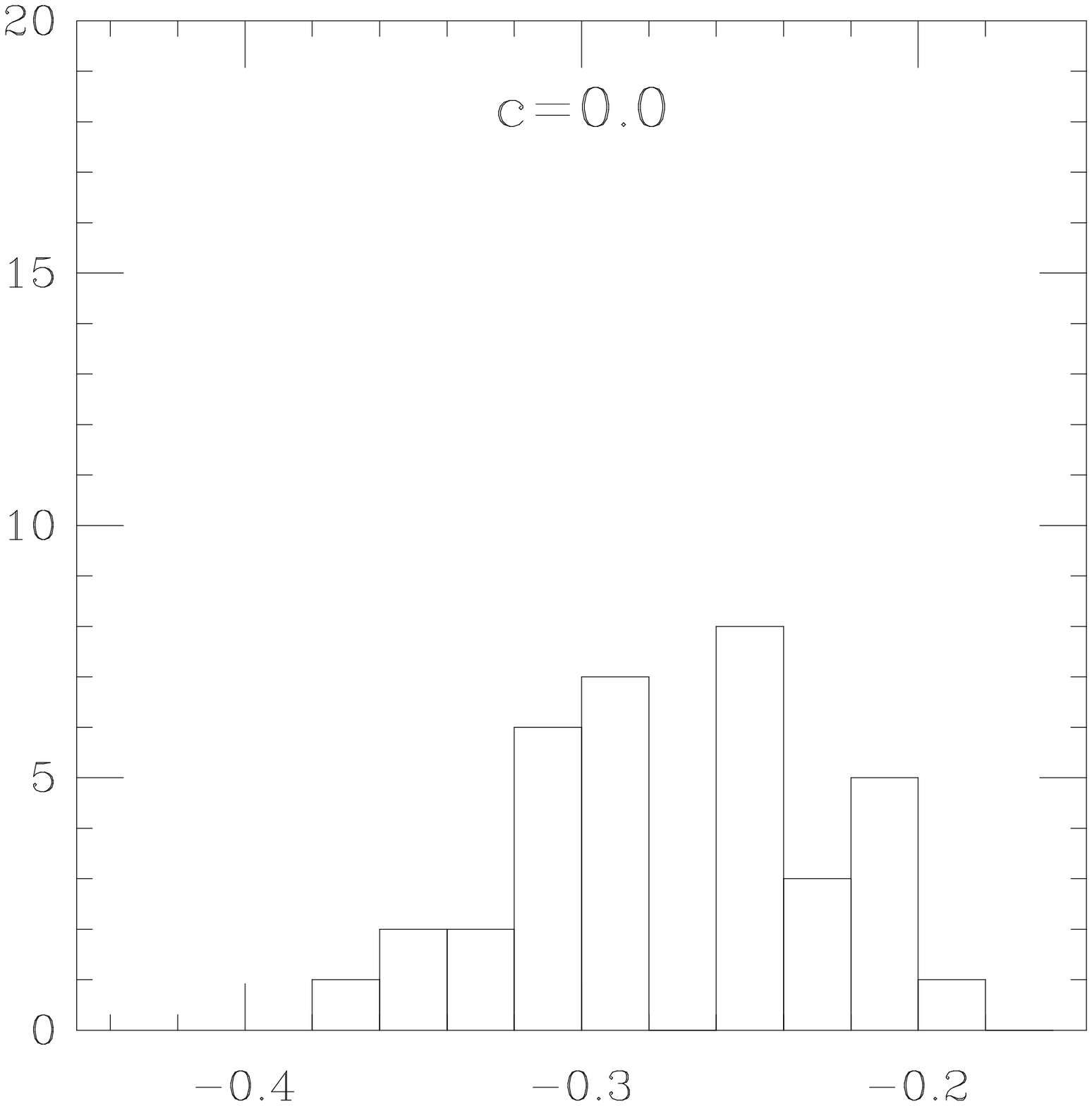}{80mm}
\ewxy{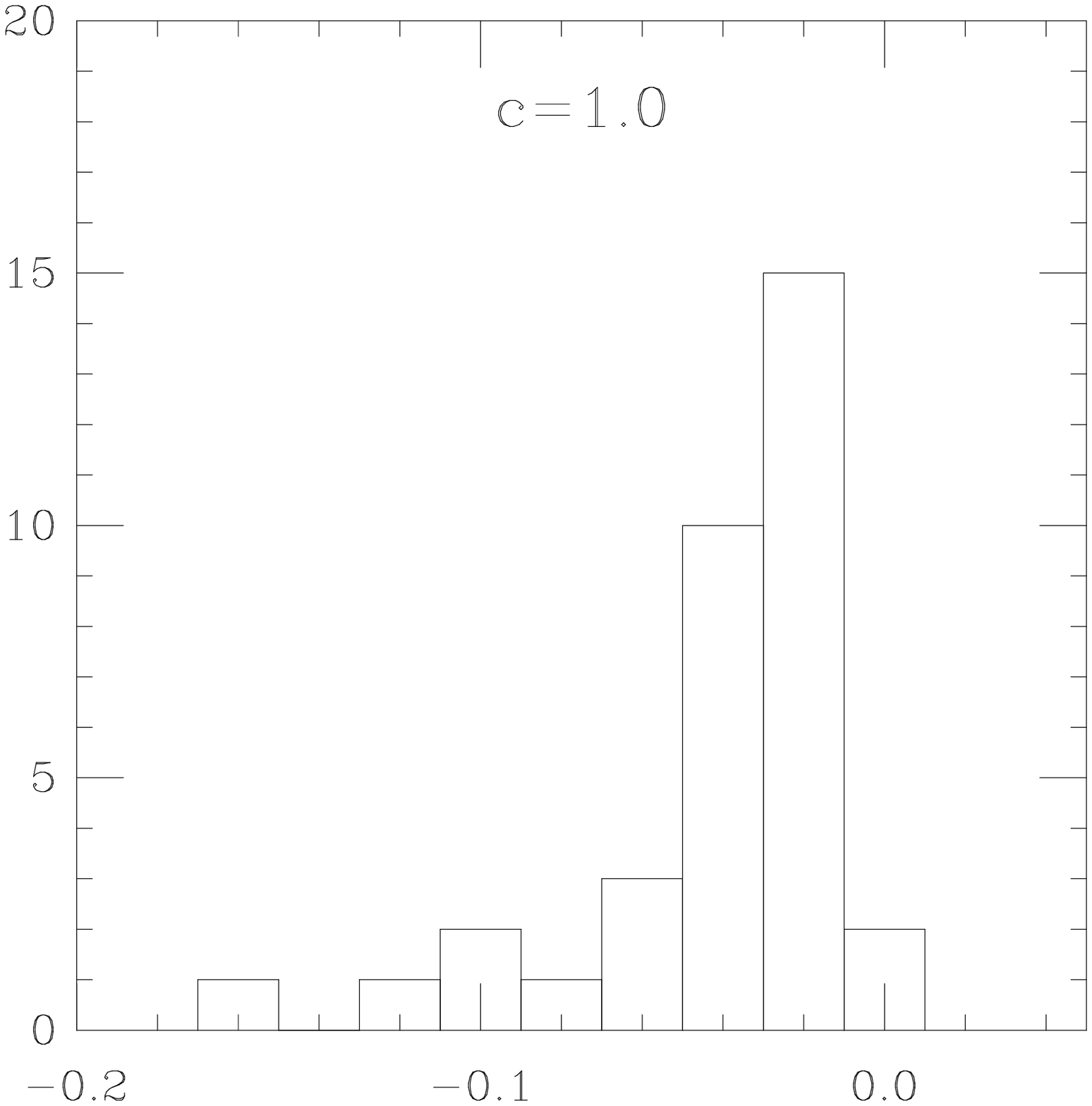}{80mm}}
\vspace{0.5cm}
\centerline{\ewxy{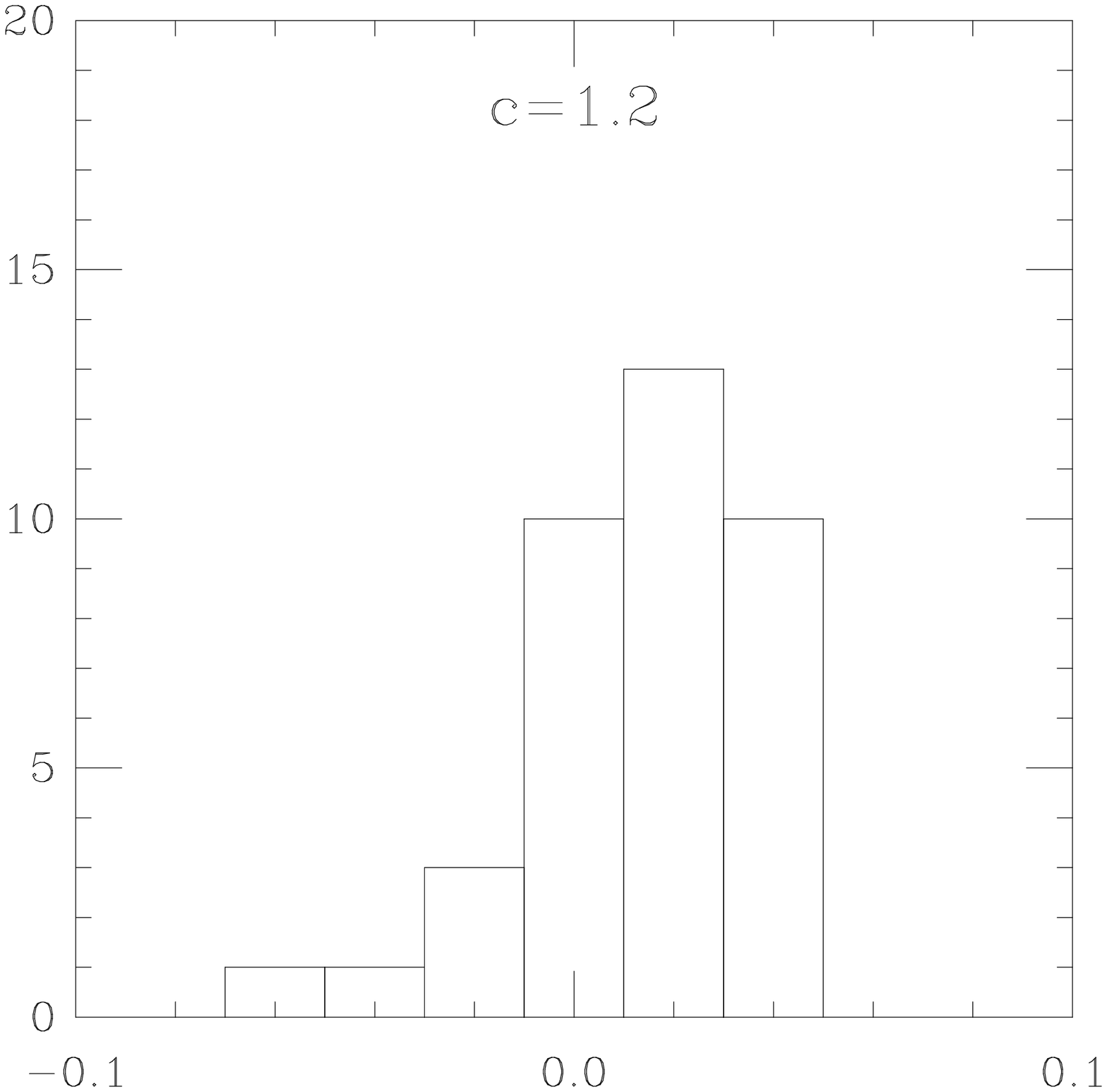}{80mm}
\ewxy{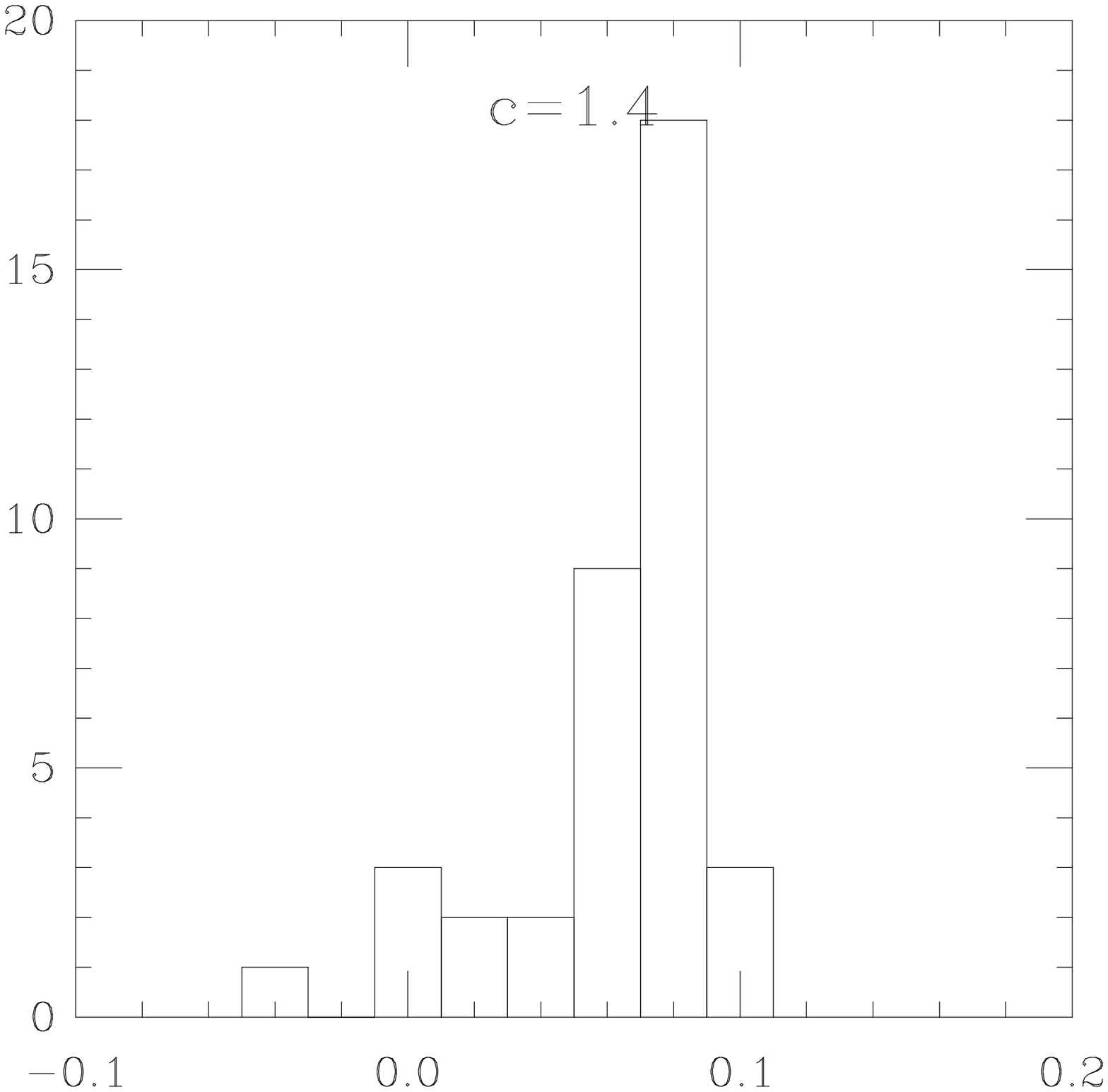}{80mm}}

\caption{Locations of the real eigenmodes at $\beta=5.8$  of fat link 
Wilson action (a); fat link clover
fermions C=1.0 (b);  $C=1.2$ (c); $C=1.4$ (d).}
\label{fig:polevsclover5.8}
\end{figure}

Figure \ref{fig:polevsclover5.8}  shows the spread of the real eigenmodes at $\beta=5.8$ as the function
of the bare quark mass for clover coefficients $C=0.0$ (Wilson action), $C=1.0$, $C=1.2$ and $C=1.4$.
 The eigenmodes
were located using the pseudoscalar density function of Eqn. \ref{PG5P} on 40 $8^4$ configurations.
The horizontal ranges of the histograms  are equal to the range the eigenmode search had been
performed. 
To the right of the distributions is the  confining phase. If one  measures the pion mass in the positive
mass region and extrapolates $m_\pi^2$ to zero with the bare mass $m_0$, 
one finds that the
bare mass at which $m_\pi=0$ lies at a value $m_c$ located within
the range of  the low-lying real eigenmodes \cite{NEW_EXCEPT}.
 If a particular configuration
has an eigenmode with $m_0 > m_c$, and if one attempts to compute
the quark propagator at $m=m_0$, the propagator will be singular.
This is (one kind of) exceptional configuration, and at a minimum
it compromises the statistical averaging process inherent in the Monte
Carlo simulation process.
To the left of the distributions, in the negative quark mass region, are the doublers. In Figure 
\ref{fig:polevsclover5.8} there is no indication of them, 
since the doublers did not show up within the range
we scanned for eigenmodes. It appears that the doublers and low lying modes are well separated at
$\beta=5.8$. This seems to be in contradiction with the statement of Ref. \cite{SCRI_spectralflow},
where the authors claim that the real modes cover the whole investigated negative quark mass range. Of
course, as
 there is a non-zero probability of finding an eigenmode anywhere, our claim 
is only that the
distribution is strongly peaked  at the low lying modes and the doublers are well separated. Since in Ref.
\cite{SCRI_spectralflow} no distribution histogram is published, we cannot tell if there is a real
discrepancy between the results or it is the question of interpreting the data. One should also note
that in \cite{SCRI_spectralflow} thin link
actions were used, so we are really discussing two different
actions. We will return to this question in \cite{NEW_EXCEPT}.

\begin{figure}
\centerline{\ewxy{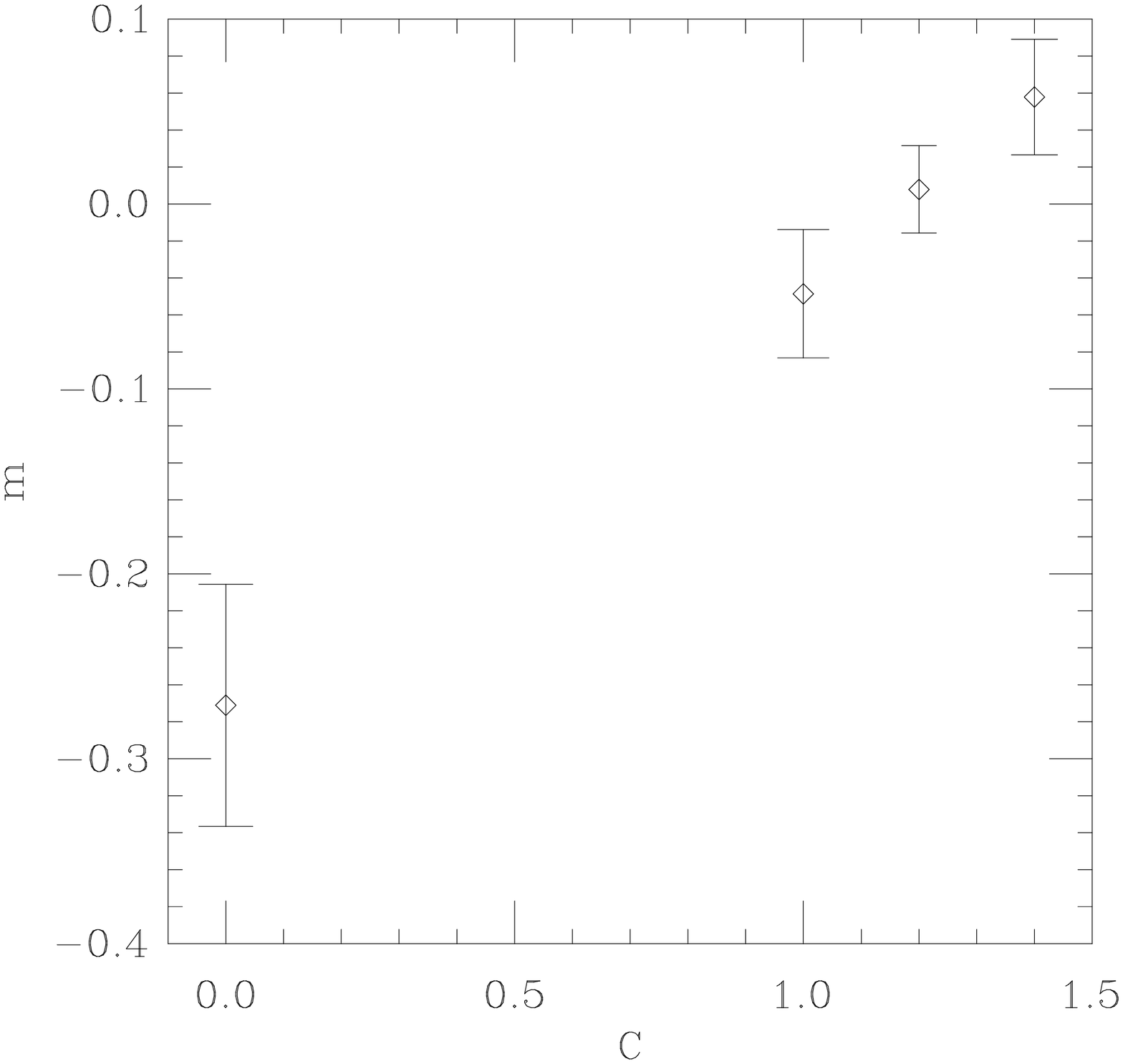}{80mm}
\ewxy{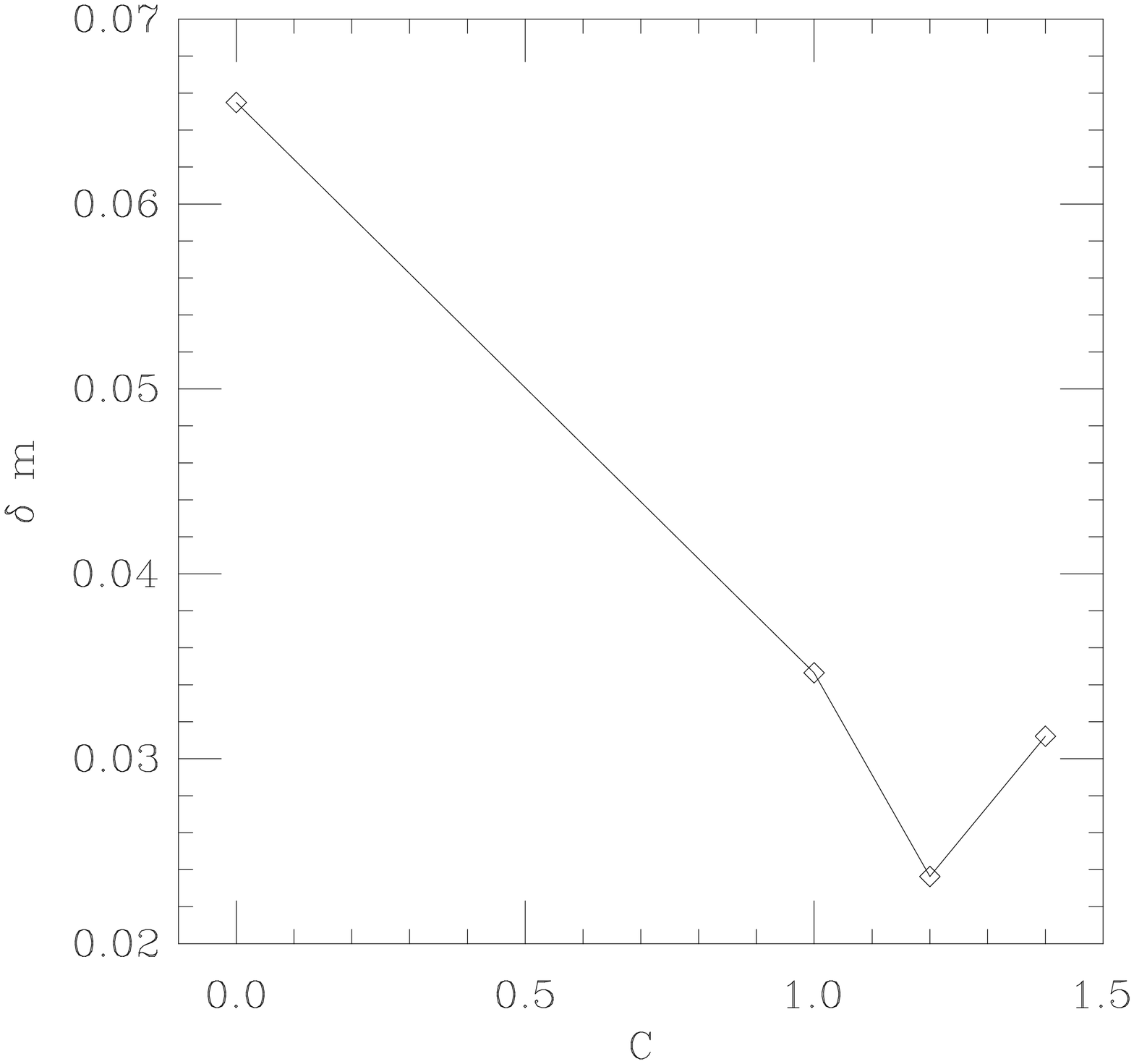}{80mm}}
\caption{(a) The average real eigenmode location as the function of the clover coefficient at $\beta=5.8$. The error
bars show the spread of the modes. (b) The spread of the modes from (a) as the function of
the clover coefficient. }
\label{fig:averpole}
\end{figure}

To quantify the spread of the low energy real modes, 
in figure \ref{fig:averpole}a we plot the average real eigenmode
location as the function of the clover coefficient. The error bars here are not errors, they are the
spread of the modes in $m$.  
Figure \ref{fig:averpole}b shows only the spread as the function of the
clover coefficient.
$m_c$, where the pion becomes massless, lies somewhere in the middle of
the mode distribution, closer to its right (large mass)
end. The average eigenmode locations
and the upper end of the error bars in figure \ref{fig:averpole}a
bracket $m_c$.
Smearing the link removes most of the additive mass renormalization even for
the Wilson action, reducing $m_c \sim -0.95$ for the thin link action 
to $m_c \sim -0.22$ for our case. 
Adding a clover term to the action further reduces the
additive mass renormalization, and even larger clover terms
induce a positive mass renormalization. At $\beta=5.8$ the additive mass 
renormalization is minimal for $c\approx 1.2$. This is also the value where the spread of the
eigenmodes is minimal.
To conclude, the results indicate that at $\beta=5.8$
 (lattice spacing $a\simeq 0.15$ fm) the low lying  and
doubler modes are well separated and explicit chiral symmetry breaking is minimized with clover
coefficient $C=1.2$ with our fat link action.

\begin{figure}
\centerline{\ewxy{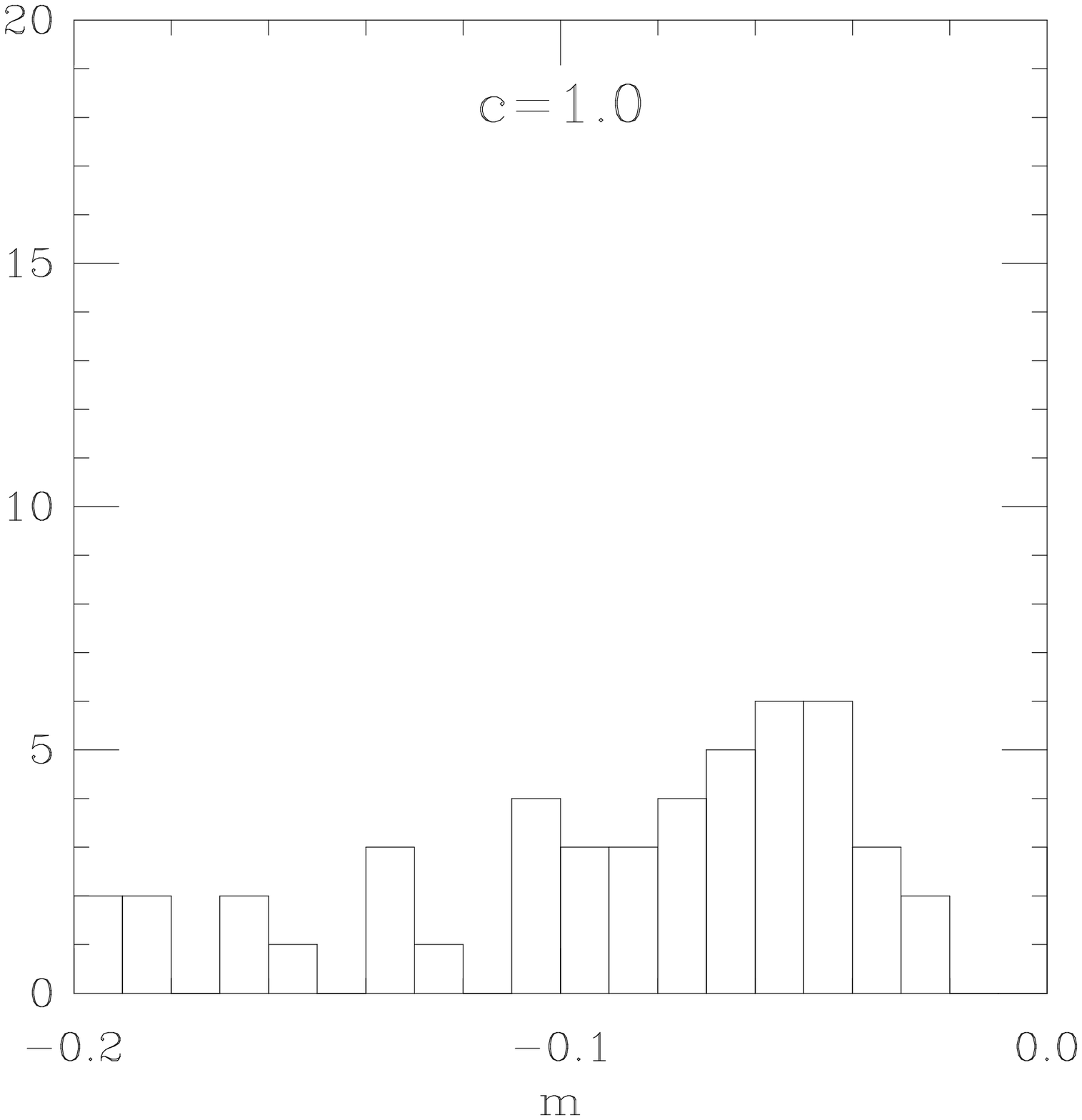}{80mm}
\ewxy{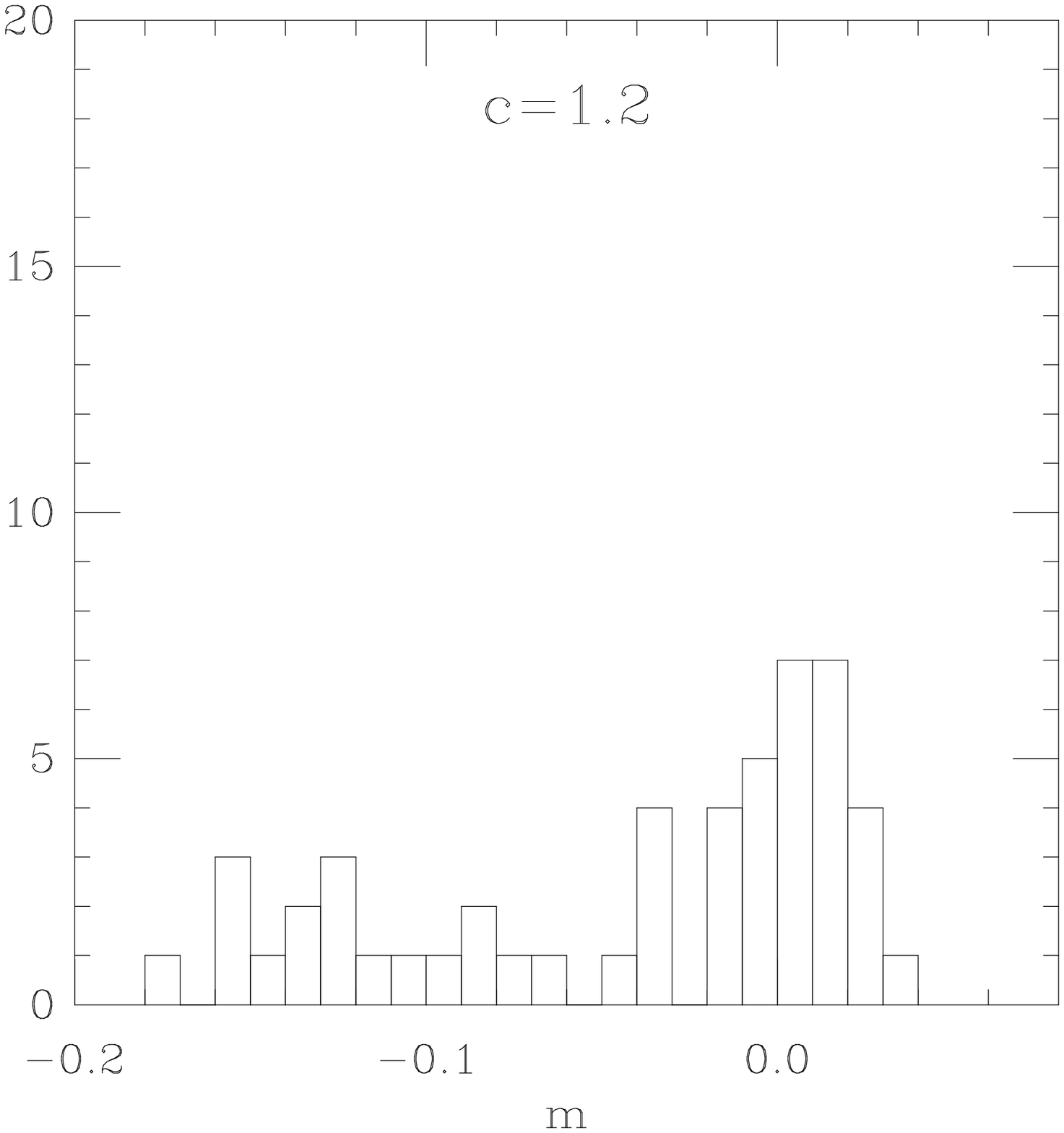}{80mm}}
\vspace{0.5cm}
\centerline{\ewxy{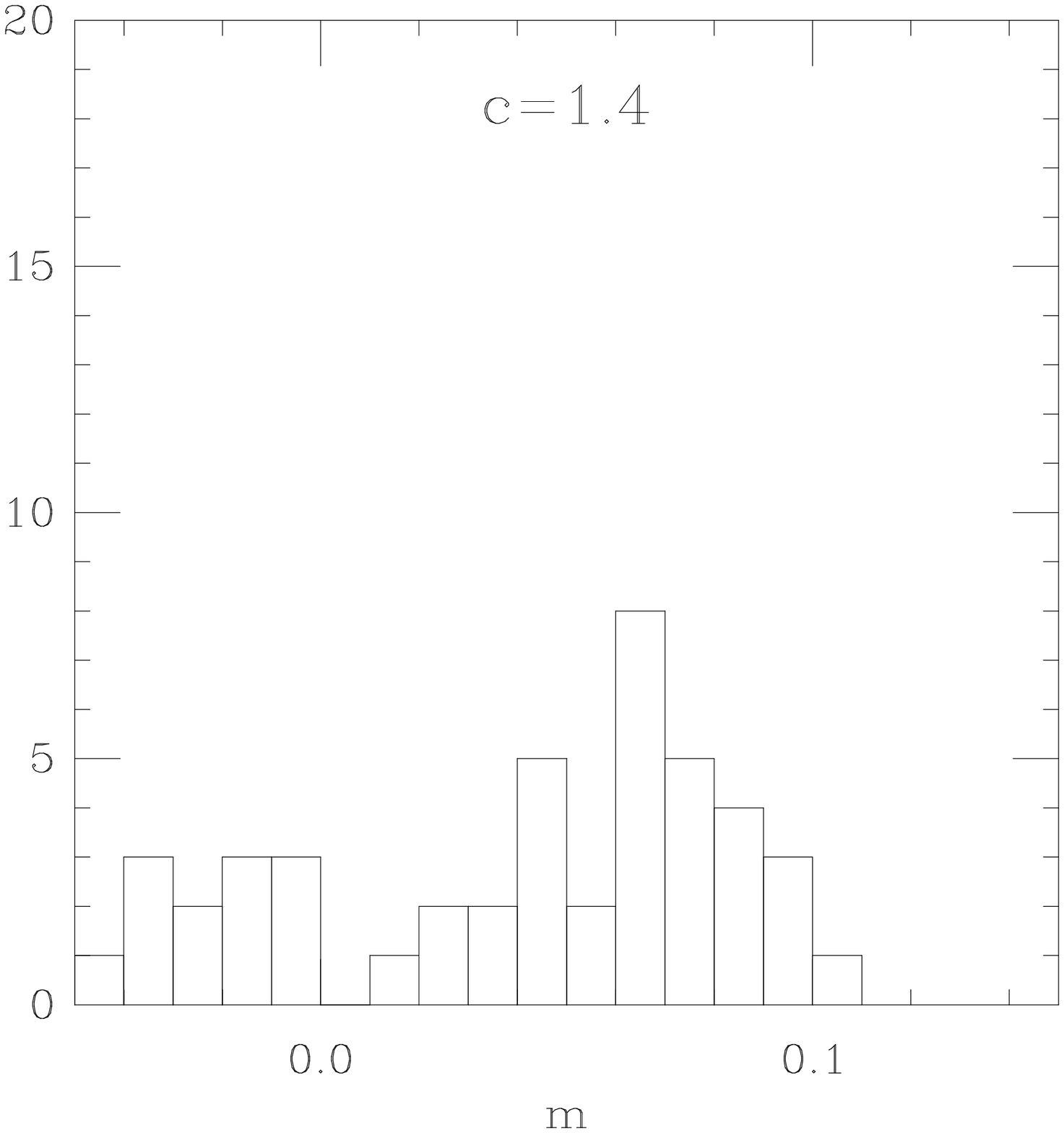}{80mm}}
\caption{Locations of the real eigenmodes at $\beta=5.7$  of fat link  clover
fermions C=1.0 (a);  $C=1.2$ (b); $C=1.4$ (c).}
\label{fig:polevsclover5.7}
\end{figure}

The situation is less convincing at $\beta=5.7$ (lattice spacing 
$a\simeq 0.20$ fm). Figure
\ref{fig:polevsclover5.7}, again based on 40 $8^4$ configurations,
 shows the distribution of the eigenmodes at $\beta=5.7$, clover coefficient
$C=1.0,1.2$ and 1.4. Here the low lying  modes are not as well separated than for $\beta=5.8$. For
$C=1.0$ the distribution has a large tail extending towards negative quark masses. The situation is a
bit
better for $C=1.2$. Since the low lying modes are not well
separated from the doublers, it makes no sense to calculate the spread of the distribution. The
continuum description of chiral symmetry breaking is about to break down at $\beta=5.7$. If any
of the distributions of figure 
\ref{fig:polevsclover5.7} describes continuum physics, it is $C=1.2$ where the
overall mass renormalization is close to zero and the low lying modes are best separated form the
doublers. 

\begin{figure}
\centerline{\ewxy{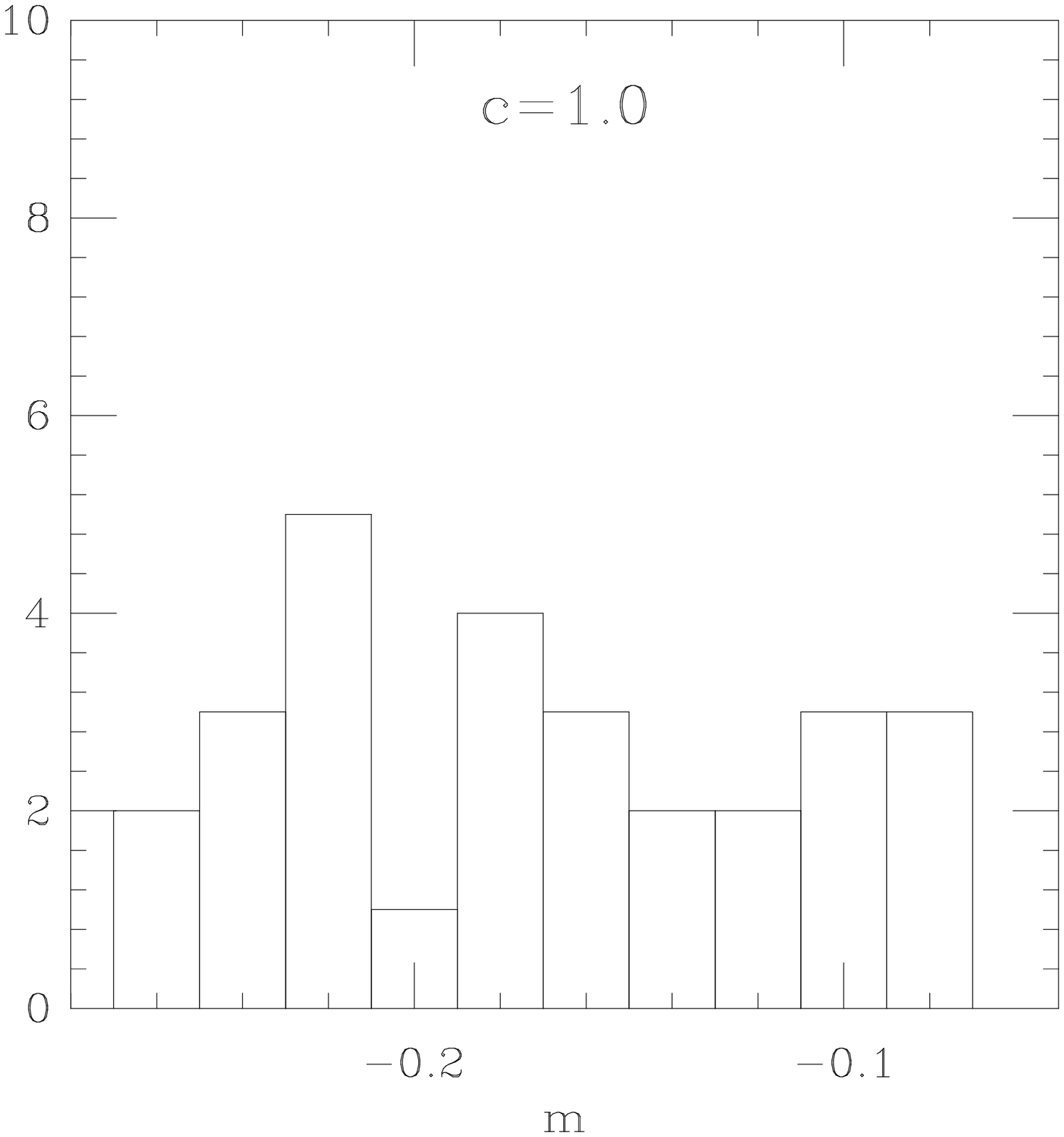}{80mm}
\ewxy{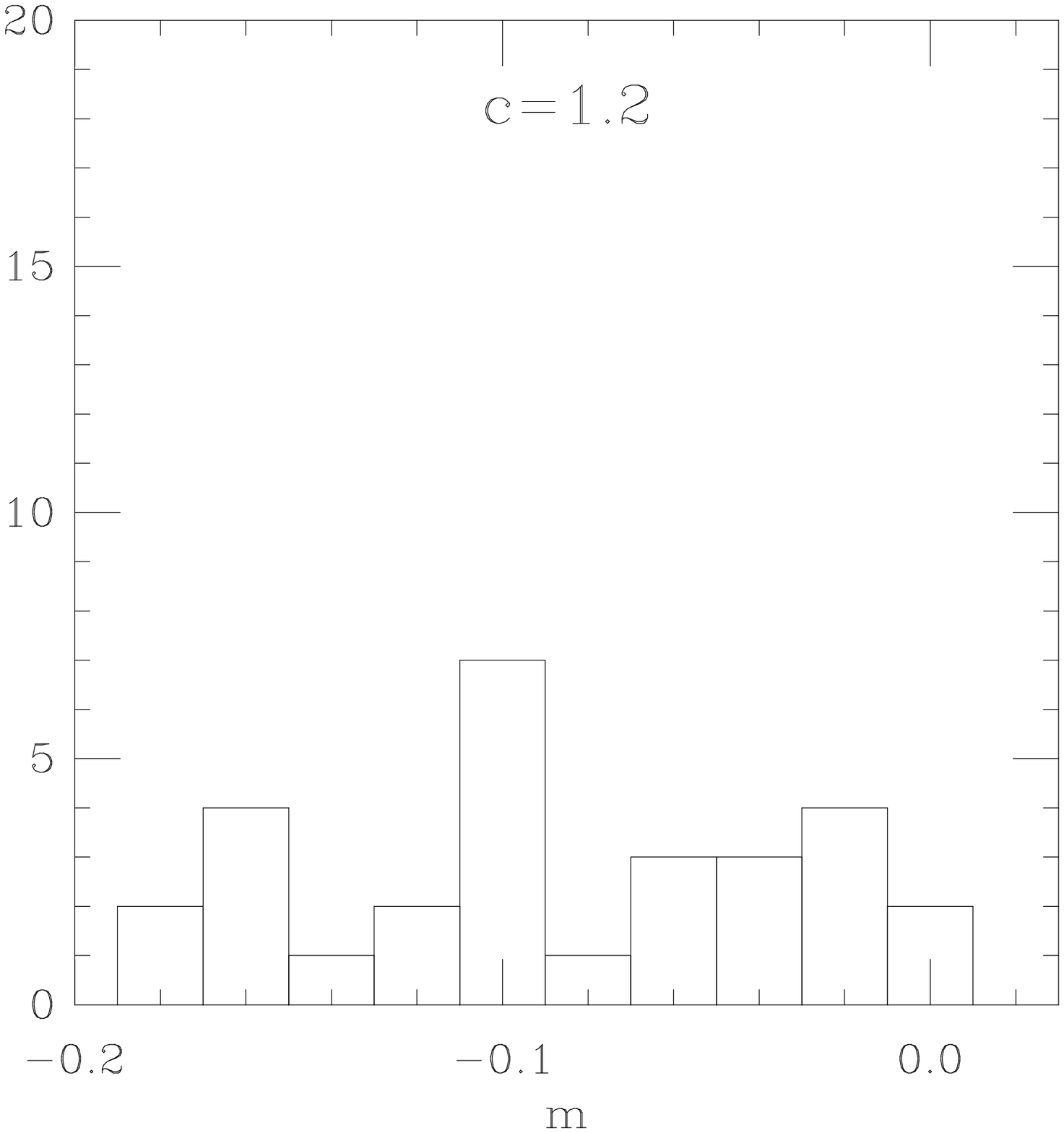}{80mm}}
\caption{Locations of the real eigenmodes at $\beta=5.55$  of fat link  clover
fermions C=1.0 (a);  $C=1.2$ (b).}
\label{fig:polevsclover5.55}
\end{figure}

Finally, at $\beta=5.55$ (lattice spacing $a\simeq 0.24$ fm)
Fig. \ref{fig:polevsclover5.55} shows that the distribution
of eigenmodes is broad and the low lying  modes and doublers are completely
mixed.

These figures show that it is not possible to make the lattice spacing
greater than about 0.2 fm, and still retain the continuum-like description
of chiral symmetry breaking using a clover-like action. Other, better tuned actions might perform
better at large lattice spacing. Our exploratory studies with a 
hypercubic fixed point \cite{HYPER} action showed
that the FP action is not better in terms of chiral symmetry breaking than the fat link clover action
with $C=1.0$. Its dispersion relation was improved
at lattice spacing $a=0.36$ fm, and it (and the clover action)
showed only a small amount of scale violation in hyperfine splittings
at large lattice spacing.
This just shows that scaling or near scaling of a few quantities does not guarantee that the
physical mechanism responsible
for  chiral symmetry breaking is the same as in the continuum.
 This should serve as a warning sign for any calculations
at large lattice spacing with actions of untested chiral properties.

In principle, one should optimize not the spread of the bare quark mass, 
but the spread of eigenmodes in terms of some physical observable, 
like the pion mass. However, at large fattening, the relation between the
bare quark mass and the pion mass shows little variation with $C$.

The usual nonperturbative tuning of the clover term \cite{LUSCHER}
attempts to optimize current algebra via PCAC relations.
It is not clear to us how this is related to what we do, since
it has been done so far only for thin link actions.  We know that
the optimal $C$ by our criterion for a thin link action is
larger than the usual nonperturbative tuning, for example 2.5 vs 2.0   at
$\beta=5.8$.  Our criterion is based on a direct attack on low energy real
eigenmodes.
To the extent that the
dynamics of QCD for light quark masses is dominated by this physics, we feel our
optimization criterion is well founded.

We do not know if there is a connection between this tuning method
and the Ginsparg-Wilson \cite{GW} realization of chiral symmetry on the
lattice. The free field limit of these fat link actions is identical
to the standard Wilson action, and so we are doing something different from
 improving the
chiral properties of the interacting theory by improving the chiral properties
of the free theory.

\section{Scaling Tests}
We have done a calculation of spectroscopy and matrix elements
at lattice spacing $aT_c=1/4$ 
($\beta=5.7$ with the Wilson gauge action)
using the fat link clover action with $C=1.2$.
This is on the edge of the allowed range of lattice spacings according
to our criterion of the last section.
By itself, one lattice spacing is not a scaling test, but we can combine
our results with those from other actions to compare the new action
to them.

\subsection{Spectroscopy}

The spectroscopy measurement is entirely straightforward. 
The lattice volume  was
$8^3\times 24$.The data set was 80 lattices.
We gauge
fixed to Coulomb gauge and used a Gaussian independent particle source
wave function $\psi(r) =
\exp(-(r/r_0)^2)$ with $r_0=2$.
We used pointlike sinks projected onto low momentum states.
We used naive currents ($\bar \psi \gamma_5 \psi$, etc.)
 for interpolating fields.
The spectra appeared to be asymptotic (as shown by good (correlated)
 fits to a single exponential)
beginning at $t\simeq 3-5$  and the best fits were selected using the 
 HEMCGC criterion \cite{HEMCGC}.

Our fiducials for comparison
are Wilson action and clover action
quenched spectroscopy.  We have tried to restrict the data
we used for comparison to lattices with the proper physical volume.

\begin{figure}
\centerline{\ewxy{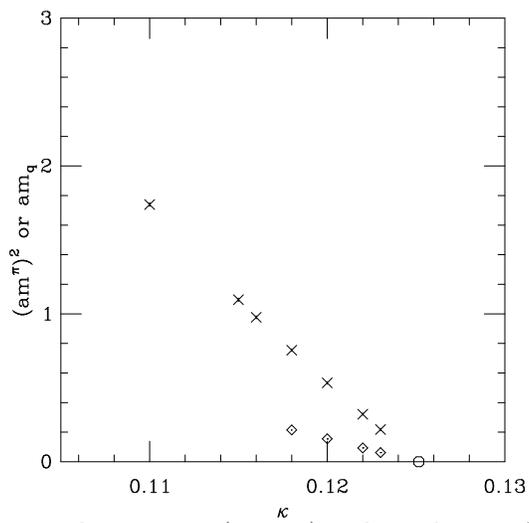}{80mm}
}
\caption{Bare squared pion mass  (crosses) and
quark mass from Eq. 12 (diamonds) 
 vs hopping parameter for the C=1.2 action.
The octagon shows $\kappa_c$.}
\label{fig:pisqc1.2}
\end{figure}

We can roughly estimate the critical bare quark
mass (at which the pion is massless)
by  linearly extrapolating $m_\pi^2$ to zero in $m_0$.
We also estimated the critical bare mass using the PCAC relation
\bee
\nabla_\mu \cdot \langle \bar\psi \gamma_5 \psi(0) \bar\psi\gamma_5 \gamma_\mu
\psi(x) \rangle =
2m_q \langle\bar\psi \gamma_5 \psi(0) \bar\psi\gamma_5 \psi(x) \rangle .
\ee
which, going to the lattice and following \cite{DOUGFPI} 
is done by fitting the pseudoscalar source-pseudoscalar sink to
\bee
P(t) = Z(\exp(-m_\pi t) + \exp(-m_\pi (N_t -t) ) )
\ee
and the pseudoscalar source-axial sink to
\bee
A(t) = {Z_P \over Z_A}
{{2m_q}\over m_\pi} Z(\exp(-(m_\pi t) - \exp(-m_\pi (N_t -t) ))
\ee
to extract $m_q$.
Fig. \ref{fig:pisqc1.2} shows the
squared pion mass vs bare quark mass. 
The quark masses from the local pseudoscalar and axial currents
are also shown.

Fits of the squared pion mass $(am_\pi)^2 = B(1/\kappa -1/\kappa_c)$
give $B=1.55(2)$ and $\kappa_c=0.12515(7)$.
The quark mass is fit to $am_q = A(1/\kappa -1/\kappa_c)$
and we find $A=0.447(7)$, $\kappa_c=0.12515(6)$. In free field theory,
we would expect $A=1/2$, $\kappa_c=1/8$,
 and this is our first hint that perturbative corrections
to bare quantities are small.

\begin{figure}
\centerline{\ewxy{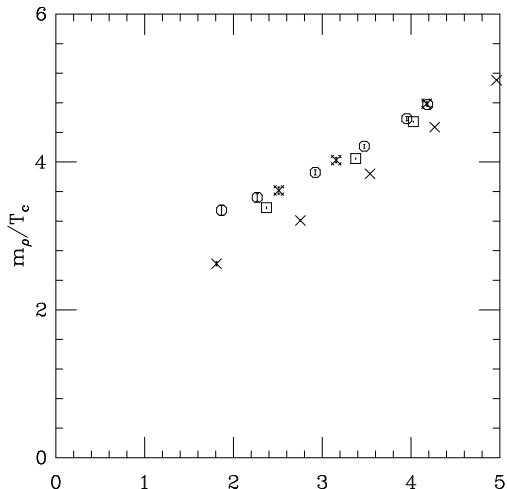}{80mm}
}
\caption{Octagons show $m_\rho/T_c$ vs. $m_\pi/T_c$ for  the C=1.2 action
at $aT_c=1/4$. Also shown are Wilson action data, with
crosses for $aT_c=1/4$,
squares for $aT_c=1/8$,
and
fancy crosses for $aT_c=1/12$.
}
\label{fig:rhotccl}
\end{figure}

\begin{figure}
\centerline{\ewxy{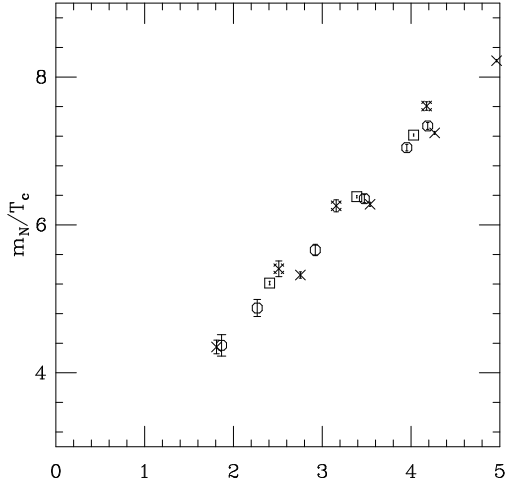}{80mm}
}
\caption{$m_N/T_c$ vs. $m_\pi/T_c$ for  the C=1.2 action
and the Wilson action, labelled as in Fig. 6.}
\label{fig:ntccl}
\end{figure}

As a scaling test we compare
$m_\rho/T_c$ and $m_N/T_c$ vs. $m_\pi/T_c$ for the C=1.2 action in
 Figs.  \ref{fig:rhotccl}
and  \ref{fig:ntccl}.

\begin{figure}
\centerline{\ewxy{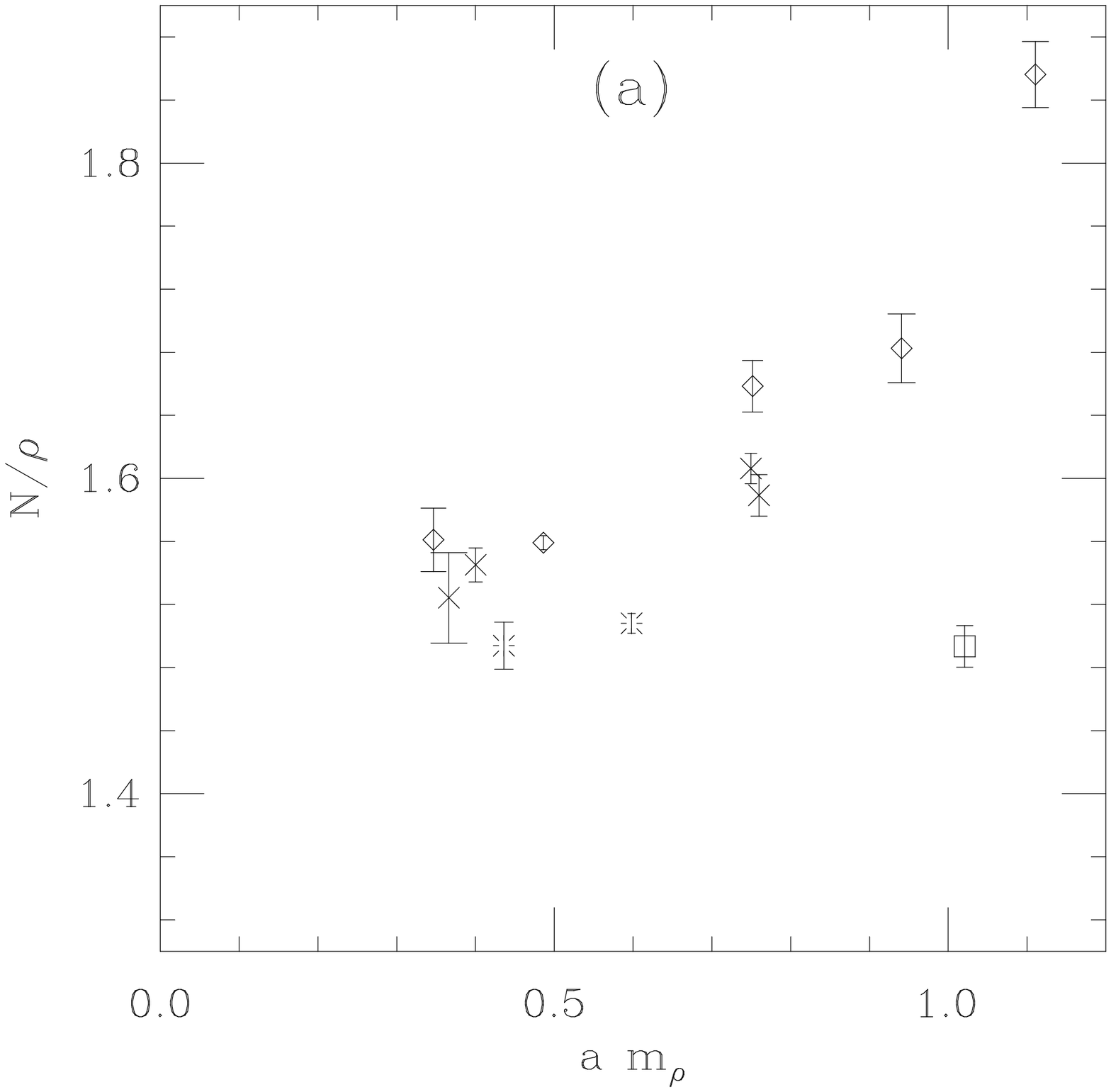}{80mm}
\ewxy{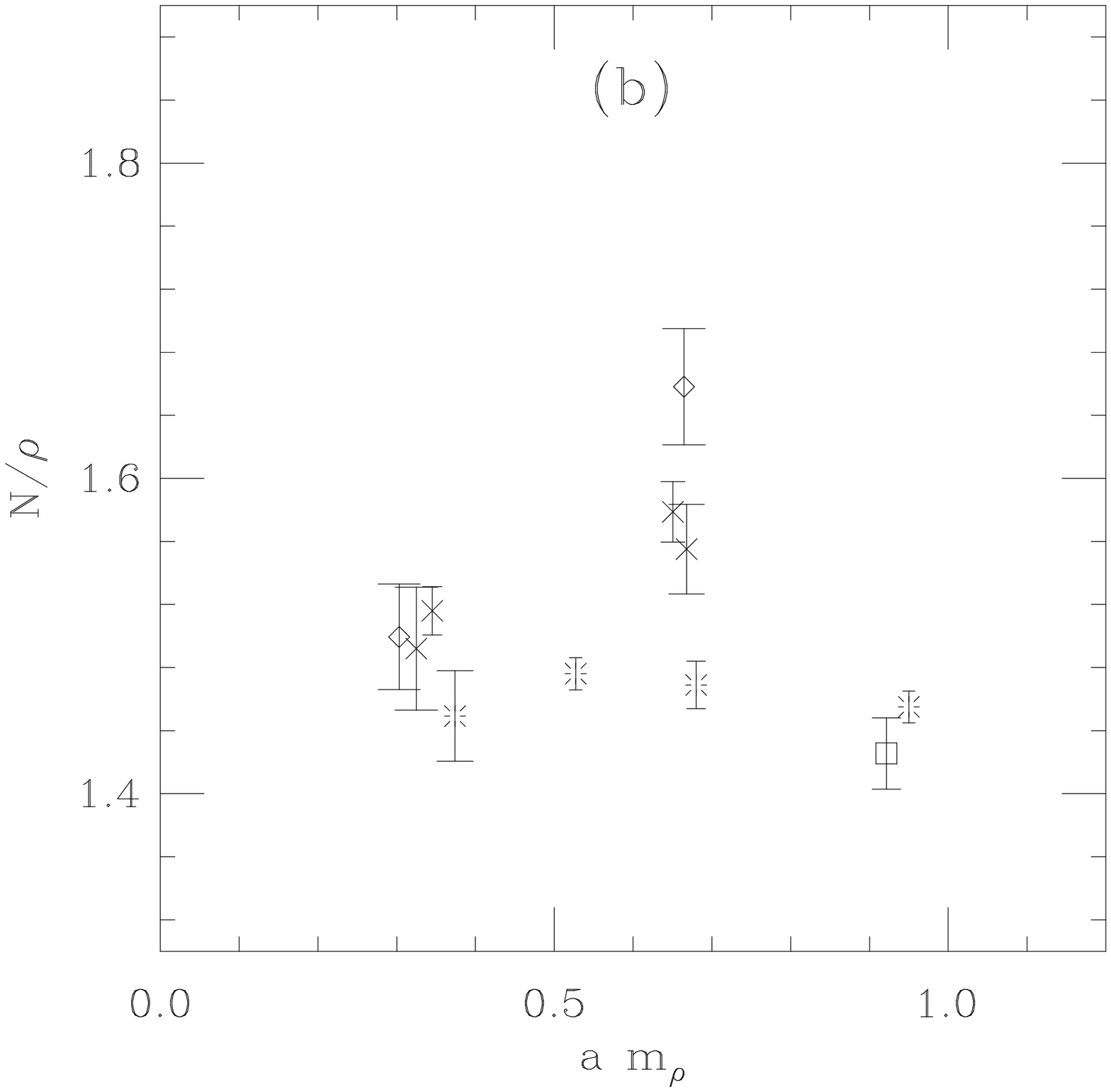}{80mm}}
\vspace{0.5cm}
\centerline{\ewxy{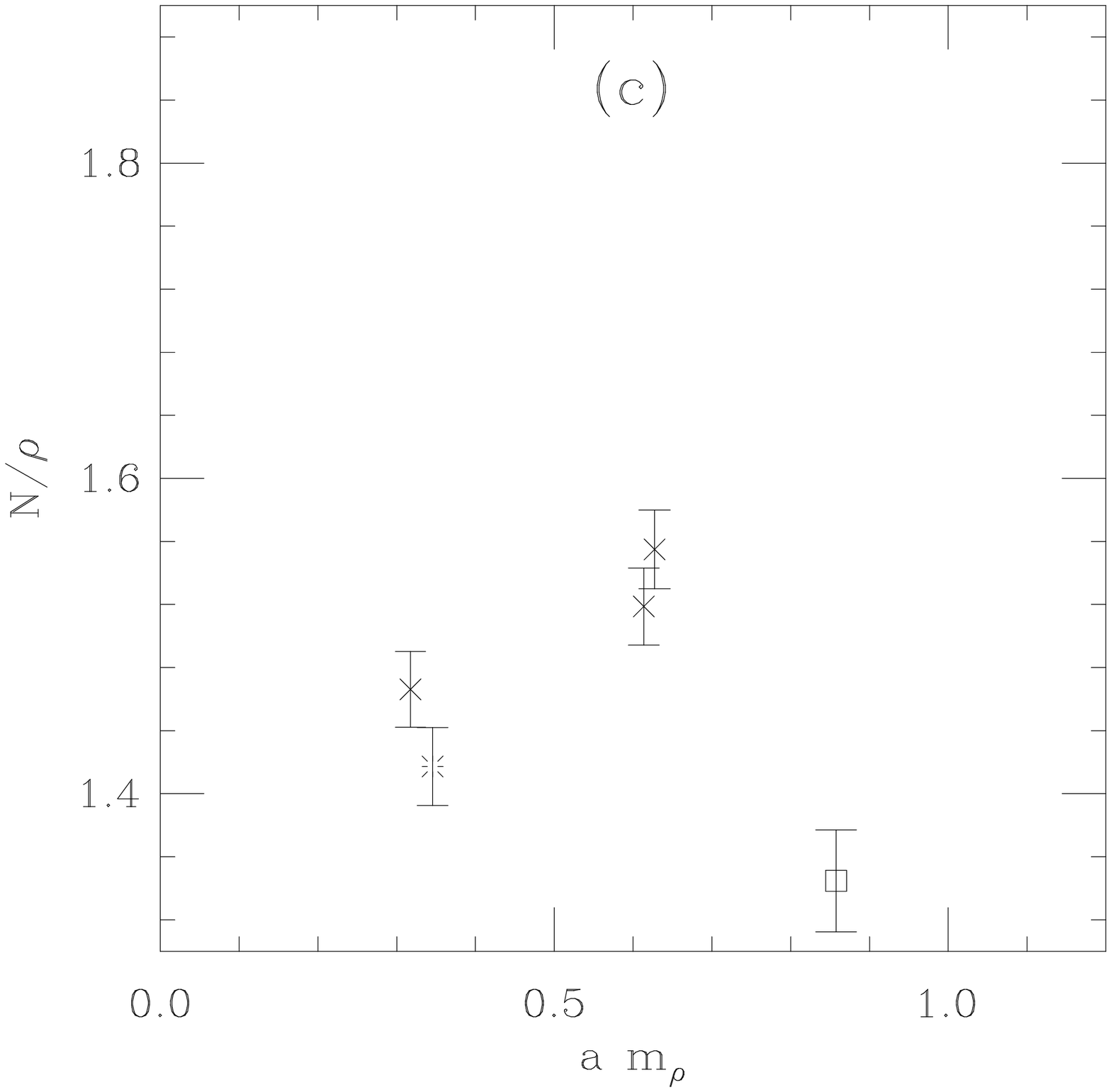}{80mm}}
\caption{A scaling test for the C=1.2 action (square)
vs. Wilson actions on lattices of fixed physical size (diamonds)
and larger volumes (crosses), and the
nonperturbatively improved clover action (bursts).
 Data are interpolated to
$\pi/\rho=0.80$ (a), 0.70 (b), and 0.60 (c).}
\label{fig:allrat}
\end{figure}

We  compare scaling violations in hyperfine splittings by
interpolating our data to fixed $\pi/\rho$ mass ratios and plotting
the $N/\rho$ mass ratio vs. $m_\rho a$. We do this at three
$\pi/\rho$ mass ratios, 0.80, 0.70 and 0.60, in Fig. \ref{fig:allrat}.
In these figures the diamonds are Wilson action data in lattices of fixed
physical size,
$8^3$ at $\beta=5.7$ \cite{BUTLER},
$16^3$ at $\beta=6.0$ \cite{DESY96173}
$24^3$ at $\beta=6.3$ \cite{APE63},
 and the crosses are data in various larger lattices:
$16^3$ and $24^3$ at $\beta=5.7$ and $32^3$ at $\beta=6.17$ \cite{BUTLER},
$24^3$ at $\beta=6.0$ \cite{DESY96173}.
When they are present the data points from larger lattices illustrate
the danger of performing scaling tests with data from different volumes.
The bursts are from the nonperturbatively improved clover action of Refs.
\cite{ALPHA} and \cite{SCRI}.
The squares show the C=1.2 action.

We give a table of masses from the C=1.2 action in Table
 \ref{tab:w570}.

We conclude that the fat link clover action has at least as good
scaling behavior as the usual nonperturbatively tuned clover action.

We also noticed that the new action seems to be more convergent than 
the thin link clover action: the same biconjugate gradient code needs about
half as many steps to converge to the same residue as the usual thin link 
clover action, for the same $\pi/\rho$ mass ratio.

We did not encounter any exceptional configurations with the fat link
action for $m_\pi/m_\rho \ge 0.56$.

\subsection{Dispersion Relations}

\begin{figure}[htb]
\begin{center}
\vskip -10mm
\leavevmode
\epsfxsize=60mm
\epsfbox[40 50 530 590]{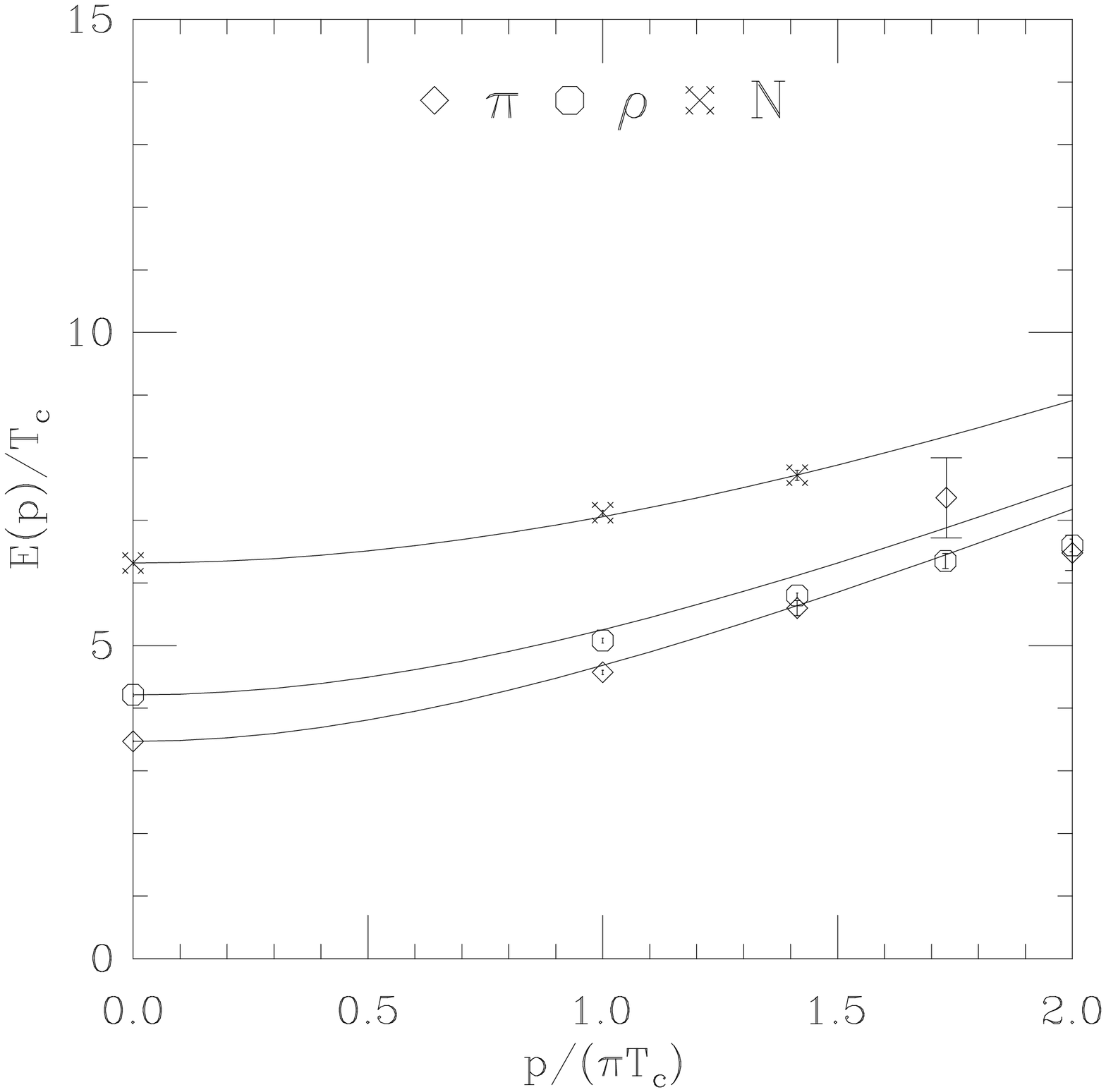}
\vskip -5mm
\end{center}
\caption{Dispersion relation for  hadrons at $aT_c=1/4$ ($a \simeq 0.18$
fm)  from the C=1.2 action.
The curves are the continuum dispersion relation for the appropriate
(measured) hadron mass.}
\label{fig:eptc}
\end{figure}

To view the dispersion relation, we first
plot $E(p)$, the energy of the state produced with spatial momentum
$\vec p$, as a function of $|\vec p|$.
The result for   the C=1.2 action  at $\kappa=0.118$ is compared to the
free dispersion relation 
at $aT_c=1/4$ in Fig. \ref{fig:eptc}.
It looks very similar to results from the Wilson action at the same
lattice spacing--an entirely unsurprising result.

\begin{figure}
\centerline{\ewxy{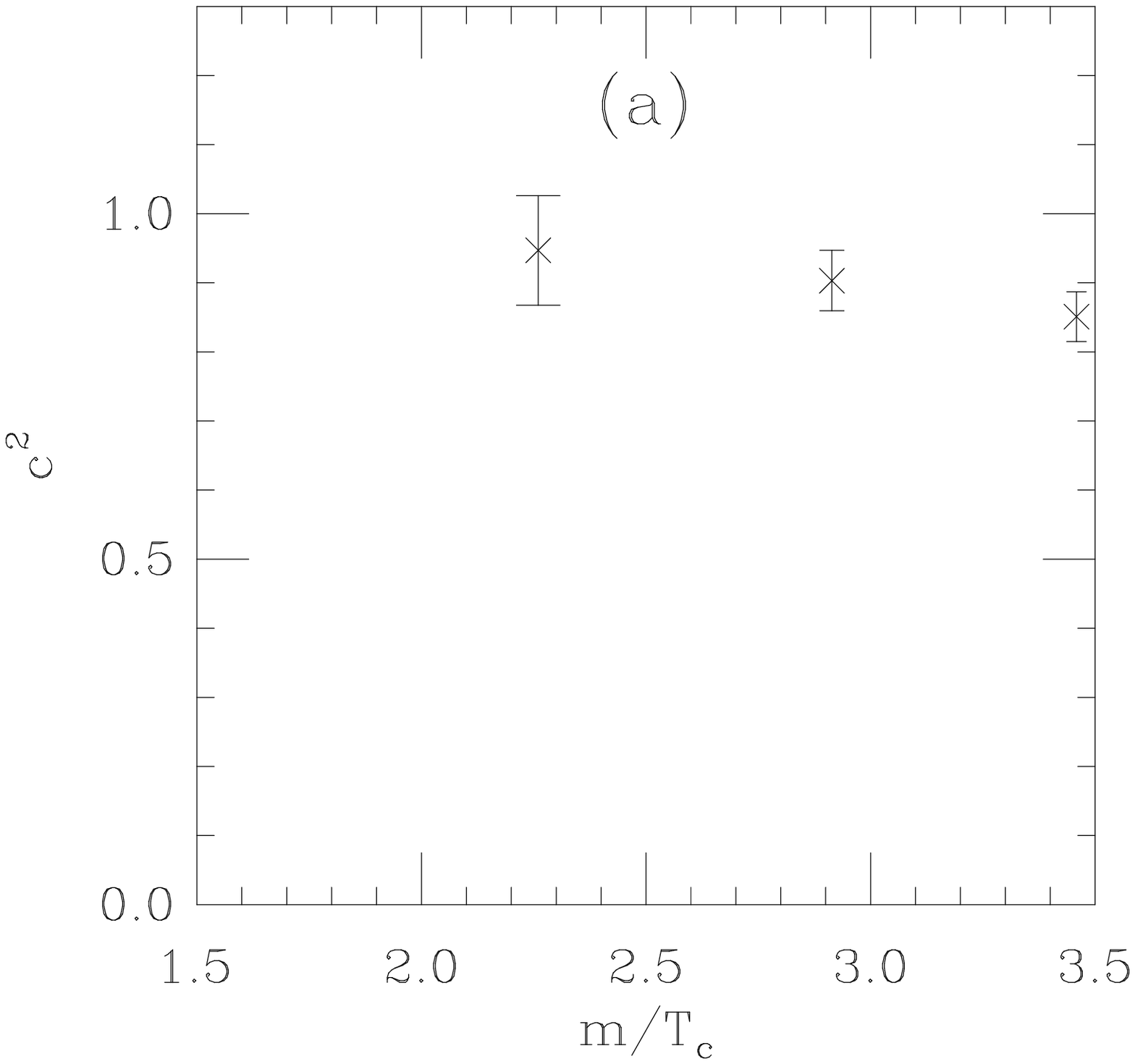}{80mm}
\ewxy{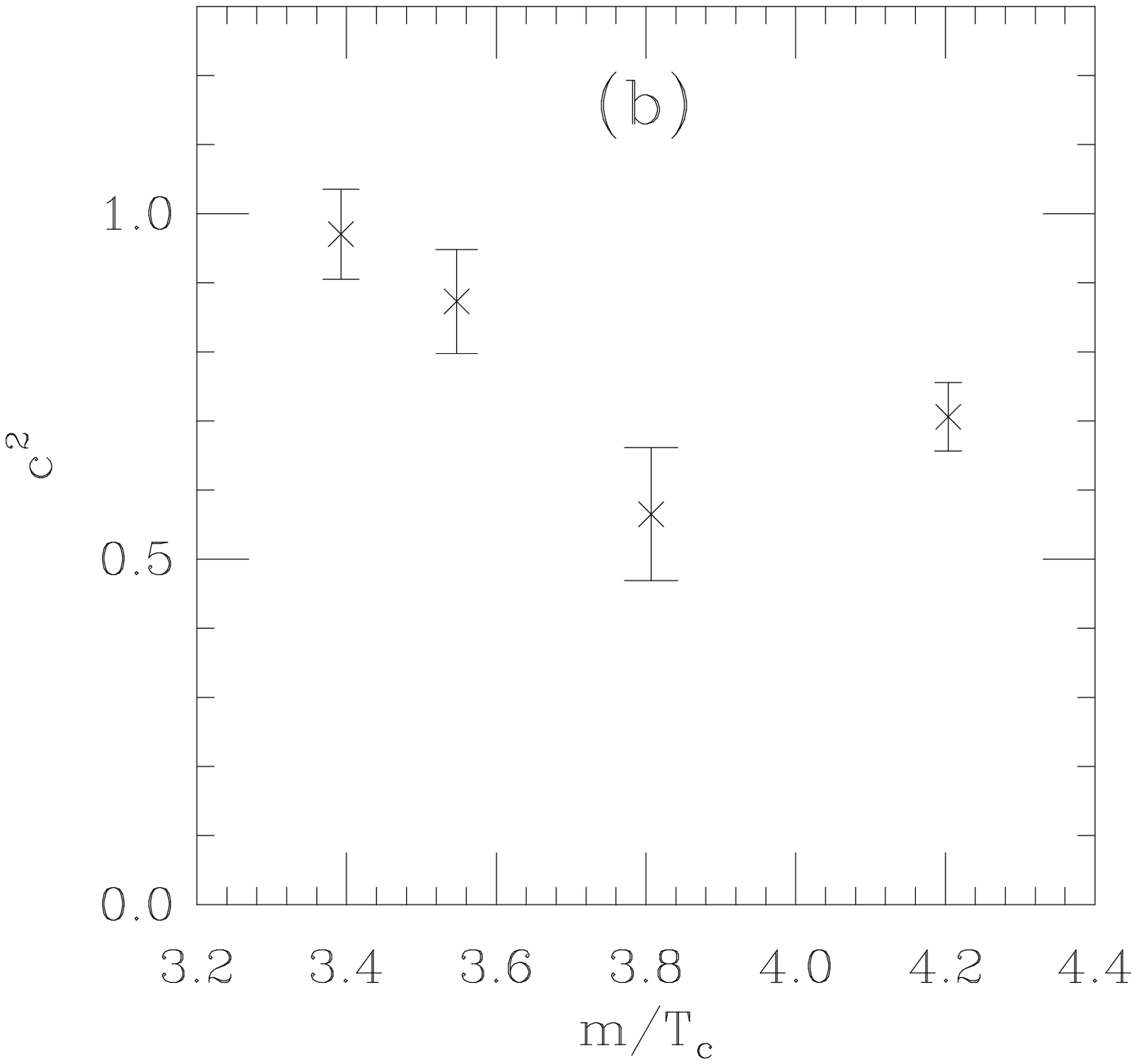}{80mm}}
\vspace{0.5cm}
\centerline{\ewxy{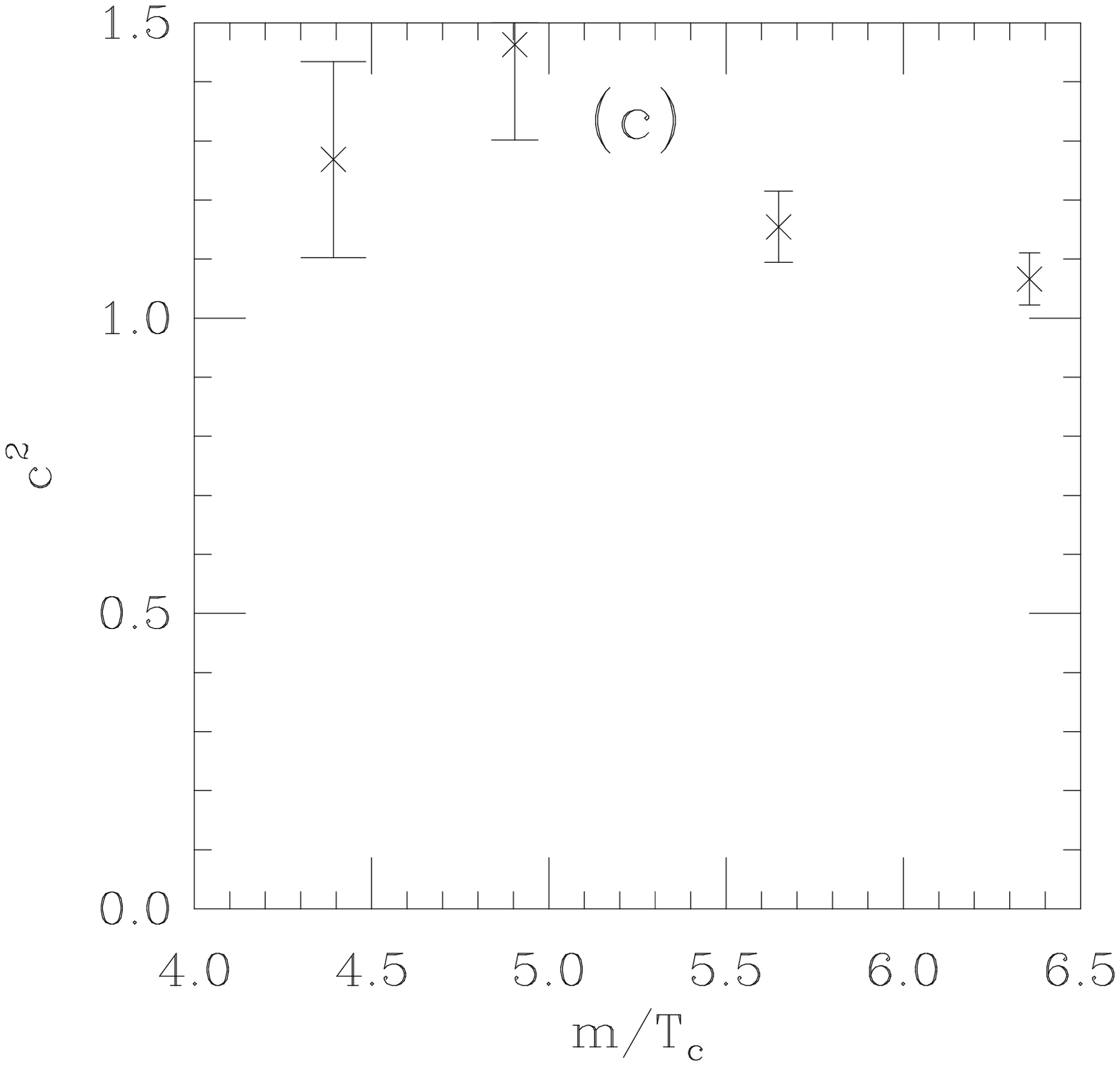}{80mm}}
\caption{Squared speed of light vs. hadron mass in units of $T_c$,
for (a) pseudoscalars, (b) vectors) and (c) protons, from the
C=1.2 action at $\beta=5.7$.}
\label{fig:csq}
\end{figure}

This behavior is quantified by measuring the squared speed of light,
$c^2 = (E(p)^2-m^2)/p^2$, for $\vec p= (1,0,0)$.  We do this by performing
a correlated fit to the two propagators. The result is presented in 
Fig. \ref{fig:csq} and shows the worsening of the dispersion relation
at larger quark mass, characteristic  Wilson/clover kinetic behavior.

\begin{table}
\begin{tabular}{|c|l|l|l|l|l|}
\hline
$\kappa$ & PS  & V  &  N  &  $\Delta$ \\
\hline
   0.118 &  0.868( 3) &  1.053( 6)&  1.589(13) &  1.725(19)  \\
   0.120 &  0.730( 4) &  0.965( 8)&  1.415(19) &  1.593(22)  \\
   0.122 &  0.567( 5) &  0.880(12)&  1.219(28) &  1.501(24)  \\
   0.123 &  0.467( 6) &  0.837(17)&  1.093(36) &  1.432(27)  \\
\hline
\end{tabular}
\caption{Table of best-fit masses, $C=1.2$, $\beta=5.7$.}
\label{tab:w570}
\end{table}

\subsection{Renormalization factors}
We now turn to a (rather naive) set of measurements of simple matrix
elements.

Formally, the fat link action can be regarded as ``just another''
 $O(a^2)$ improved action, if one assumes that the coefficient of
the clover term is actually $C= 1+ O(g^2)$.  In perturbation theory,
which involves the vector potential $A_\mu$ rather than the link,
the action shows the usual cancellation between the clover term
and the scalar $\bar \psi \psi A$ term. The spectrum is $O(a^2)$;
matrix elements using ``rotated fields'' $\psi \rightarrow
(1 -(\gamma\cdot D )/2) \psi$ are $O(a^2)$ improved.  Of course, the 
lattice-to-continuum 
renormalization Z-factors
are different than the usual thin link clover Z-factors.

In principle, one could calculate all 
 Z-factors using lattice perturbation theory. The Feynman
rules differ from the rules for the usual clover action, only in
 that the vertices are multiplied by form factors, which are
functions of the gluonic momentum.  We have not tried to do this yet.
Since we are working at fairly strong coupling, we felt that it would
be easier to compute the Z-factors nonperturbatively, beginning with
the vector current and then computing the axial 
renormalization factor using Ward identities.  Since this is just the
first calculation using this action, we will restrict ourselves to
the naive (local) currents.

We begin with the vector current.  The conserved (Noether) current
for the fat link action is the same as the Wilson current, just with
fat links instead of thin links,
\bee 
J_\mu^{cons}(n) = {1 \over 2} (\bar \psi(n)(\gamma_\mu -1)V_\mu(n)\psi(n+\hat \mu)
- \bar \psi(n+\hat \mu)(\gamma_\mu +1)V_\mu^\dagger(n)\psi(n) ).
\ee
This current is conserved but not improved.  We choose to define the 
vector current Z-factor from the ratio of  forward matrix elements
\bee
Z_V = {{\langle \pi(T) \pi(0) \rangle} \over
 {\langle \pi(T) J_0(t)\pi(0) \rangle}}
\label{ZV}
\ee
for $0<t<T$.
The local current is defined as $J_\mu^{loc}(n) = \bar \psi(n) \gamma_\mu \psi(n)$.
We measured the Z-factors using 20 configurations on a periodic
 $8^3 \times 24$
lattice with $T=10$ and averaged $9>t>1$. The results of a jacknife analysis
are shown in Table \ref{tab:zv}.

The fact that $Z^{cons} \ne 1$ is a finite size effect.  In the numerator
of Eqn. \ref{ZV} both quarks can propagate in  both time directions around
the torus, but the forwards and backwards paths contribute differently
to the denominator. We correct this phenomenologically by computing the
ratio $Z^{loc}/Z^{cons}$ and display it in the table.

Lattice perturbation theory which correctly takes into account the residue
of the pole of the massive quark\cite{ZFACTOR}  predicts that this ratio is
equal to 
\bee
Z_V= {{1-6 \kappa}\over{2\kappa}} \hat Z_V
\ee
 where the $(1-6 \kappa)$ factor is the
residue (recall $\kappa_c=.125$) and $\hat Z_V = 1 + a g^2 +\dots$. 
 As Table \ref{tab:zv} shows, the correction to the tree level 
formula, $\hat Z$, which is what is usually quoted as the Z-factor,
differs from unity by less than  than two per cent at $\kappa=0.123$.

We also attempted to measure the axial vector current renormalization
using the chiral Ward identity \cite{CHIRALWI}.
With 50 lattices, we only had a signal at $\kappa=0.118$, where
we found $Z_A=1.210(8)$.  This is again quite close to the pure
kinetic result of 1.237.  Dividing out the kinematic factors,
this would give $\hat Z_A=0.978$.

At $\beta=6.0$, Ref. \cite{MPSV} found $\hat Z_V=0.824$, $\hat Z_A=1.09$ for
the usual clover action, and improved operators.

\begin{table}
\begin{tabular}{|c|l|l|l|l|l|}
\hline
$\kappa$ & $Z^{cons}$  & $Z^{loc}$  &  $Z^{loc}/Z^{con}$  & 
${{1-6\kappa}\over{2\kappa}}$ \\
\hline
   0.118 &   1.020(1) & 1.253(2) & 1.228 & 1.237 \\
   0.120 &   1.032(2) & 1.185(4) & 1.148 & 1.166 \\
   0.122 &   1.054(5) & 1.129(6) & 1.071 & 1.098 \\
   0.123 &  1.071(6) & 1.127(8) & 1.052 & 1.065  \\
\hline
\end{tabular}
\caption{Vector current renormalization factors, $C=1.2$, $\beta=5.7$.}
\label{tab:zv}
\end{table}

 It would be very interesting to measure the mixing
of left- and right-handed operators (for $B_K$) in this action.

\section{Conclusions}
We have shown that via a combination of fattening the links and
tuning the magnitude of the clover term,
 it is possible to optimize the chiral behavior of
the clover lattice fermion action.  The example we presented
reduces the spread in the low energy real eigenmodes by about a factor of three
in units of the squared pion mass, compared to the usual Wilson action.
We fixed the fattening and varied the clover coefficient, but it is clear
that a real optimization would involve varying both factors.

The same procedure can be applied to the usual clover action, without
fat links, but  the value of the clover term $C$ which minimizes the
spread of real eigenmodes due to instantons is so large that  new
kinds of configurations, whose singular behavior is unrelated to
fermion modes sitting on individual instantons, become exceptional
\cite{NEW_EXCEPT}.
Thus this choice of action  does not represent a good choice for 
improvement.

The method of construction exposes an apparent fundamental upper limit
for the lattice spacing in a QCD simulation of about  0.2 fm. At larger
lattice spacing, the doubler modes and the low energy modes do not show
a clean separation, and the mechanism of chiral symmetry breaking 
is qualitatively different than in the continuum.
The Wilson and clover actions, and the
 one improved action we tested, the hypercubic action of Ref. \cite{HYPER},
failed at  lattice spacing  0.24 fm.
This test does not apply to actions involving heavy quarks, for which chiral
symmetry is (presumably) not important, but before any other improved action
can be said to reproduce continuum light hadron
physics at some lattice spacing,
it should show a separation between the spectrum of
its near low energy modes and its doubler modes.

The ``fat clover'' action we have presented in this paper, as an example
 of an action with improved chiral properties, has many other nice 
properties as well.  It appears to exhibit scaling of hyperfine
splittings at 0.2 fm. The action is quite insensitive to the UV
behavior of the underlying gauge field.
It has Z-factors for simple matrix elements which
are very close to unity. 
And finally, the amount of resources required to construct propagators
is halved compared to the usual clover action, at equivalent parameter
values.  It would be trivial to modify any existing clover code to
improve it in the way we have described.  Of course, its kinetic
properties, including the dispersion relation,  artifacts
at large $am_q$, and power law scaling violations in matrix elements,
are unchanged from the standard clover action's.

\section*{Acknowledgements}
We  would like to thank   the Colorado high energy experimental
groups for allowing us to use their work stations.
Part of the computing was done on the 
Origin 2000 at the University of California, Santa Barbara.
This work was supported by the U.S. Department of 
Energy.

\newcommand{\PL}[3]{{Phys. Lett.} {\bf #1} {(19#2)} #3}
\newcommand{\PR}[3]{{Phys. Rev.} {\bf #1} {(19#2)}  #3}
\newcommand{\NP}[3]{{Nucl. Phys.} {\bf #1} {(19#2)} #3}
\newcommand{\PRL}[3]{{Phys. Rev. Lett.} {\bf #1} {(19#2)} #3}
\newcommand{\PREPC}[3]{{Phys. Rep.} {\bf #1} {(19#2)}  #3}
\newcommand{\ZPHYS}[3]{{Z. Phys.} {\bf #1} {(19#2)} #3}
\newcommand{\ANN}[3]{{Ann. Phys. (N.Y.)} {\bf #1} {(19#2)} #3}
\newcommand{\HELV}[3]{{Helv. Phys. Acta} {\bf #1} {(19#2)} #3}
\newcommand{\NC}[3]{{Nuovo Cim.} {\bf #1} {(19#2)} #3}
\newcommand{\CMP}[3]{{Comm. Math. Phys.} {\bf #1} {(19#2)} #3}
\newcommand{\REVMP}[3]{{Rev. Mod. Phys.} {\bf #1} {(19#2)} #3}
\newcommand{\ADD}[3]{{\hspace{.1truecm}}{\bf #1} {(19#2)} #3}
\newcommand{\PA}[3] {{Physica} {\bf #1} {(19#2)} #3}
\newcommand{\JE}[3] {{JETP} {\bf #1} {(19#2)} #3}
\newcommand{\FS}[3] {{Nucl. Phys.} {\bf #1}{[FS#2]} {(19#2)} #3}

\end{document}